\UseRawInputEncoding
\documentclass{article}

\usepackage[numbers]{natbib}
\usepackage{notoccite} 

\usepackage{lineno}
\usepackage{float} 
\usepackage{amssymb}
\modulolinenumbers[5]

\usepackage{amsmath}
\usepackage[ruled, vlined, lined, linesnumbered, commentsnumbered]{algorithm2e} 
\usepackage{multirow}
\usepackage{adjustbox}
\usepackage{color}
\usepackage{amsmath}
\usepackage{dirtytalk}
\usepackage{mathtools}
\usepackage{textcomp}
\usepackage[unicode]{hyperref}

\newcommand{\rom}[1]{%
  \textup{\uppercase\expandafter{\romannumeral#1}}%
}
\usepackage{authblk}

\usepackage[figuresleft]{rotating}
\usepackage{caption}
\modulolinenumbers[5]

\usepackage[margin=1.5in]{geometry}

\newbox{\myorcidaffilbox}
\sbox{\myorcidaffilbox}{\large\includegraphics[height=1.7ex]{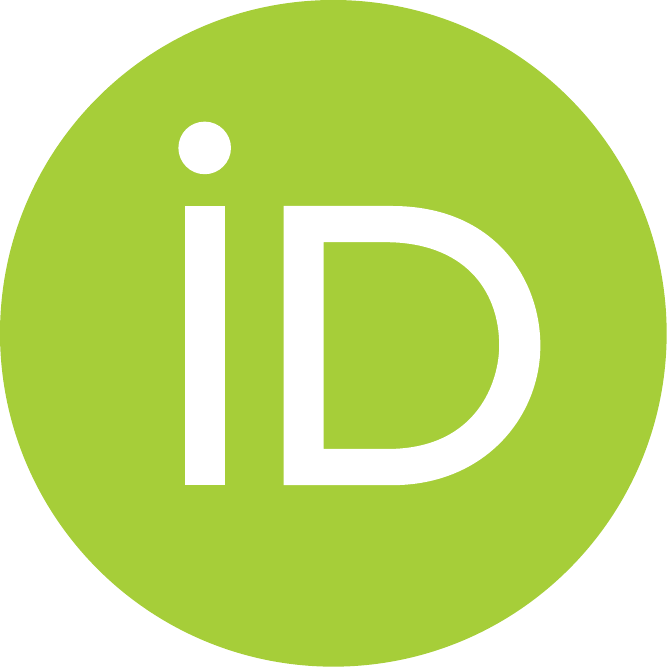}}
\newcommand{\orcidaffil}[1]{%
  \href{https://orcid.org/#1}{\usebox{\myorcidaffilbox}}}
  
\begin{document}
\title{Spherical Cap Harmonic Analysis (SCHA) for Characterising the Morphology of Rough Surface Patches\\~\\\small{[Preprint]}}
\author[1]{Mahmoud Shaqfa \orcidaffil{0000-0002-0136-2391}}
\author[2]{Gary P. T. Choi \orcidaffil{0000-0001-5407-9111}}
\author[3]{Katrin Beyer \orcidaffil{0000-0002-6883-5157}}
\affil[1,3]{\'Ecole Polytechnique F\'ed\'erale de Lausanne EPFL, Earthquake Engineering and Structural Dynamics EESD,
GCB2484 (B\^atiment GC) - Station 18 - CH-1015 Lausanne, Switzerland}
\affil[2]{Department of Mathematics, Massachusetts Institute of Technology MIT, Cambridge, MA, USA}

\date{}
\maketitle
\begin{abstract}
We use spherical cap harmonic (SCH) basis functions to analyse and reconstruct the morphology of scanned genus-0 rough surface patches with open edges. We first develop a novel one-to-one conformal mapping algorithm with minimal area distortion for parameterising a surface onto a polar spherical cap with a prescribed half angle. We then show that as a generalisation of the hemispherical harmonic analysis, the SCH analysis provides the most added value for small half angles, i.e., for nominally flat surfaces where the distortion introduced by the parameterisation algorithm is smaller when the surface is projected onto a spherical cap with a small half angle than onto a hemisphere. From the power spectral analysis of the expanded SCH coefficients, we estimate a direction-independent Hurst exponent. We also estimate the wavelengths associated with the orders of the SCH basis functions from the dimensions of the first degree ellipsoidal cap. By windowing the spectral domain, we limit the bandwidth of wavelengths included in a reconstructed surface geometry. This bandlimiting can be used for modifying surfaces, such as for generating finite or discrete element meshes for contact problems. The codes and data developed in this paper are made available under the GNU LGPLv2.1 license.
\end{abstract}

{\bf Keywords:} Spherical cap harmonics, surface parameterisation, finite element, microstructures, fractal surfaces, self-affine

\section{Introduction}
\label{SEC:INTRO}
\par
The mechanics of granular materials, such as soil mechanics, investigates the interaction of particles at various loading conditions where the shape, size and material properties of the particles can vary widely. Numerical models are therefore often used to study how the mechanical behaviour of a material is influenced by the various particle morphology and microstructures created by these particles. Similar problems exist in structural engineering when modelling, for example, the microstructure of concrete aggregates or the various stones in stone masonry elements, as the shape and roughness of the particles affect the matrices of these materials.
\par
\begin{figure}
\centering
\includegraphics[width=0.5\textwidth]{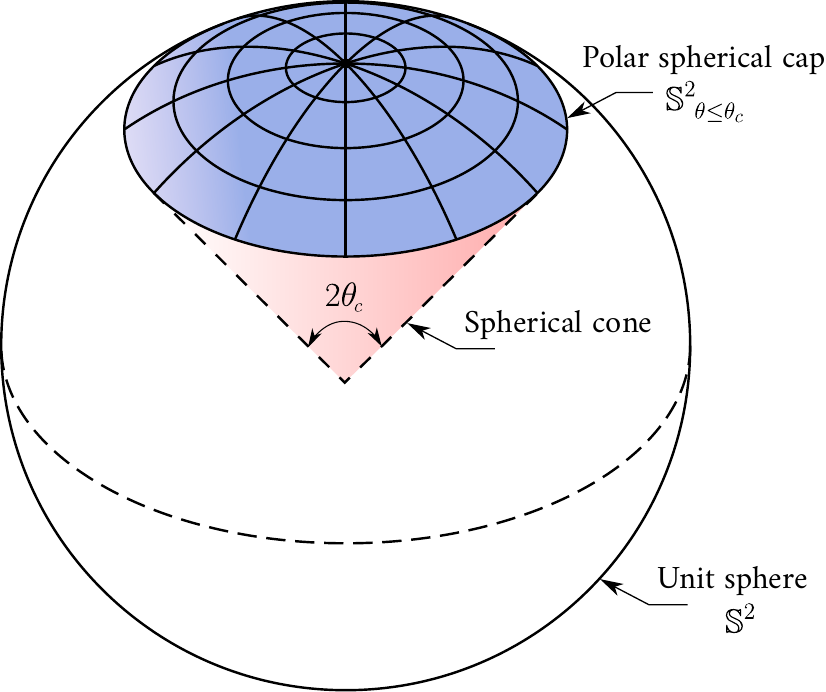}%
\caption{Representation of a polar spherical cap $\mathbb{S}^2_{~\theta \leq \theta_c}$ with half-angle $\theta_c$, where the spherical cap harmonics analysis (SCHA) is defined over.
\label{FIG:spherical_cap}
}
\end{figure}

Addressing these problems numerically requires the modelling of randomly shaped closed objects and their interactions. In such models, the shape of the particle is typically represented by a finite or discrete element mesh, while the roughness of the surface is modelled through laws that describe the interaction of two surfaces, such as friction models. Efficiently separating \say{shape} and \say{roughness} of a randomly shaped particle is a challenging problem, though work in this area helps us better understand the associated constitutive laws in numerical modelling. In recent decades, the concept of spherical harmonics (SH) analysis has been frequently used to describe the shape and roughness of stones or aggregates.
However, while this approach is transformative for closed objects, it is difficult to apply to the morphology of nominally flat and open surfaces (topological disks with free edges). In this paper, we therefore propose the use of spherical cap harmonic analysis (SCHA) for describing the geometry of open shapes. The difference here is that while a traditional spherical harmonics analysis (SHA) describes a function mapped onto an entire unit sphere, the SCHA describes a function mapped onto a spherical cap via a half-angle $\theta_c$ (see Figure~\ref{FIG:spherical_cap}). SCHs are widely used in geophysics to describe field data on planetary caps, but to our knowledge have never been used to describe the geometry of a patch with free edges. Here we give example problems where SCHA can be useful for describing the topological characterisation of a patch (i.e. an open shape with a free edge):
\begin{itemize}
    \item For studying the topology of a rough surface patch, such as a rock surface, and quantifying its fractal dimension (FD).
    
    \item For characterising different roughness regions on rocks or faults, such as sedimentary rocks made of several strata. The traditional SH approach, which assumes a closed object, only exploits basis functions to fit all the surface variations and does not distinguish regions of different roughness (multifractal surfaces).
    
    \item For matching objects based on surface characteristics. This can be useful for medical and forensic anthropology applications, such as matching fractured bones or objects (see \cite{REF:35} and \cite{REF:36} as examples), or for mechanical applications, such as characterising the surfaces of fragmented rocks or soil particles \cite{REF:33, REF:34} by directly comparing and matching the scanned fractured regions.
\end{itemize}
\par
In this paper, we propose a new method for characterising the morphology of open surface patches using SCH, focusing specifically on nominally flat patches, such as when the global radius of the curvature of the targeted patches approaches infinity. The herein proposed SCHA method allows us to study the structure of any arbitrary surface patch with faster convergence than the traditional SHA as it scales to any level of detail with reasonable computational complexity. It can also be used for reconstructing surfaces and modifying the morphology of digital twins of real surfaces, which is useful when numerically studying contact problems such as friction. Additionally, the chosen orthogonal basis functions allow us to conduct power spectral analyses of the coefficients obtained from SCHA to estimate the fractal dimensions of the surfaces.
\par
To adapt the SCHA for these applications, a new conformal parameterisation algorithm with minimal area distortion is herein proposed for parameterising the surface patches over a unit spherical cap. The parameterisation assures a unique assignment for each vertex and a uniform distribution of the morphological features over the spherical cap. Initially, the algorithm parameterises the surface patch to a planar disk using the conformal mapping algorithm proposed by Choi and Lui (2015) \cite{REF:2} followed by a suitable rescaling. Then, using the south-pole inverse stereographic projection, we conformally map the disk to a polar spherical cap. Finally, we find an optimal M\"obius transformation to minimise the area distortion on the conformal spherical cap.

\par
This paper is structured as follows: We first provide a brief literature review in Section~\ref{SEC:LITREVIEW}. We then introduce the SCHA basis functions and their solutions in Section~\ref{SEC:SCH}. In Section~\ref{SEC:PARAMETRIZATION}, we describe the proposed conformal mapping algorithm and then demonstrate the overall SCHA process in Section~\ref{SEC:SCHA}. Next, we explain the shape descriptors and the derived fractal dimensions (FD) in Section~\ref{SEC:FRACT}. We solve the problem of finding the optimal $\theta_c$ for the analysis in Section~\ref{SEC:HALF_ANGLE_PATCH}. In Section~\ref{SEC:ROUGHPROJ}, we explain how to modify the morphology by a proposed roughness projection methodology for the microstructures. The complete results are depicted in Section~\ref{SEC:NUM_EXP}. Finally, we summarise the main results of the paper and discuss possible future works in Sections~\ref{SEC:CON} and~\ref{SEC:FUTURE}, respectively.

\section{Literature review}
\label{SEC:LITREVIEW}
\par
We herein briefly present key publications on SHA and its application in mechanics, generation and reconstruction methods for both closed objects and those with free edges as well as SCHA and its current applications.

\subsection{Spherical harmonics analysis (SHA) and its applications in mechanics}
\label{SUBSEC:LIT_SPH}
\par
The modern spherical harmonics analysis (SHA or SPHARM) for closed and nonconvex particles was developed by Brechb\"uhler and K\"uble (1995) \cite{REF:1}, who applied it to the field of medical imaging for brain morphometry. SHA describes surfaces with few shape descriptors that are invariant when rotating, translating and scaling the shapes. To apply SHA to any closed, nonconvex surface, Brechb\"uhler et al. (1995) proposed a parameterisation method that provides a unique mapping of each vertex on a surface onto a unit sphere $\mathbb{S}^2$ through the use of equivalent coordinates $x(\theta, \phi)$, $y(\theta, \phi)$ and $z(\theta, \phi)$. In other words, their method can describe 3D, closed and randomly-shaped objects with a relatively small number of shape descriptors derived from 2D signals mapped onto a unit sphere. The SHA method can therefore be considered a 3D generalisation of the traditional Elliptic Fourier Descriptors (EFDs) method (see \cite{REF:3, REF:4, REF:5}), which was developed to model 2D shapes (1D signals). The SHA method can also be used to accurately reconstruct 3D shapes from their shape descriptors. Recently, Su et al. \cite{REF:71} found a correlation between the SHA and EFD descriptors and used that for predicting the size of the irregular particles from 2D images.
\par
Since the original SHA work was published, many contributions generalised it for a broader range of applications. While the original SHA dealt with voxel-based and simply connected meshes, the {Control of Area and Length Distortions} (CALD) algorithm \cite{REF:10} extended the parameterisation method to triangulated meshes, such as Standard Tessellation Language (STL) meshes. The original SPHARM determines the coefficients of the matrix by a least-squares algorithm that requires the calculation of the inverse. Modelling complex shapes with small details requires such a vast number of vertices that calculating the inverse is burdensome with the traditional SHA approach. Shen and Chung (2006) \cite{REF:8} used an iterative approach to expand the coefficients for the complex and large-scale modelling of shapes. {Other notable works include the weighted SPHARM method by Chung et al. (2007)~\cite{REF:11}, the SPHARM-PDM statistical toolbox for shape descriptors by Styner et al. (2006)~\cite{REF:6}, and the 3D analytic framework with improved spherical parameterisation and SPHARM registration algorithms developed by Shen et al. (2009) \cite{REF:9}.}
\par
In biomedical imaging, shape descriptors based on SHA were used to reconstruct surfaces and compare the morphology of organs. In engineering mechanics, two further questions were investigated: (i)~How can SHA shape descriptors be linked to the well-known, traditional shape descriptors? (ii)~How can the SHA shape descriptors be linked to mechanical parameters, such as friction coefficients, and traditional morphology measures, such as fractal dimensions (FD)? Here, we review studies in mechanics that used SHA for generating particles as well as for simulating particle mechanics.
\par
Zhou et al. (2015) \cite{REF:12} scanned real particles, computed the distribution of the SHA shape descriptors and then used these descriptors as inputs for a particle generator. The particle generator used the shape descriptors to reconstruct particle shapes, which became inputs for Discrete Element Method (DEM) simulations. Wei et al. (2018a) \cite{REF:13} established links between the coefficients expanded from the SHA and traditional shape descriptors, namely, the form, roundness and compactness, and used these as inputs for a particle generator. They applied their method to Leighton Buzzard sand particles, which were scanned by Zhao et al. (2017)~\cite{REF:14} to reveal an exponential decay of the rotation-invariant descriptors against the degree of expansion on a log-log scale. In Wei et al. (2018b) \cite{REF:15}, the authors estimated the Hurst exponent (H) from the exponential decay of the power spectrum computed from SHA coefficients; thus the SHA is able to capture a self-similar phenomenon akin to the traditional Power Spectral Density (PSD) analysis associated with Fourier expansion for 1D signals. Apart from methods based on power spectral analysis, the authors in \cite{REF:69, REF:68} correlated the FD in SHA by investigating the surface area evolution at an increasing reconstruction degree in a log-log scale.
\par
More recently, Wei et al. (2020) \cite{REF:16} studied the contact between rough spheres and a smooth base using finite element analysis. For this, they constructed roughened spheres using spherical harmonics and considering more than $2000$ degrees. The locally roughened patch on the sphere was generated by assuming a certain FD and relative roughness measure. In contrast with the traditional concept of computing Fourier coefficients along certain lines (see Russ [1994] \cite{REF:17} for details), the advantage of SHA-based fractal dimensions is that they are not direction-dependent because they are intrinsically defined everywhere along the sphere and do not expand sampled sections (1D signals). Furthermore, the SH basis functions constitute a complete orthogonal set integrated over the entire surface of the sphere in Hilbert space $\mathbb{S}^2$, which allows the power spectral analysis. Lately, Feng (2021) \cite{FENG2021113750} proposed a new energy-conserving contact model for triangulated star-shaped particles reconstructed with the SH. The model reduces the computational complexity of the discrete element simulations with controllable mesh size obtained via the golden spiral lattice algorithm.
\par
The SHA is therefore a very powerful tool for contact analysis. However, it is currently only valid for closed objects. In this paper, we address this by offering an alternative method for analysing the surface morphology---instead of analysing the surface of an entire closed object using SHA, we study the morphology of a patch on this object using SCHA. Depending on the size of the patch on a sphere, this usually requires significantly less degrees. In addition, this method makes it possible to analyse regions of different roughness on the same closed object. Here, we provide a general procedure suitable for nominally flat and open surface patches and further show how the fractal dimension concept is an intrinsic property in such basis functions defined over spheres, hemispheres and spherical caps.

\subsection{Modelling arbitrarily shaped objects with free edges}
\label{SUBSEC:REVIEW_OPENSURF}
\par
For modelling arbitrarily shaped objects using SHA, most of the literature relates exclusively to genus-0 closed objects. SHA requires an appropriate parameterisation algorithm (bijective function) over a unit sphere $\mathbb{S}^2$, but does not cover open objects, which have a free boundary along their edge. This problem was previously addressed using a hemispherical harmonic (HSH) basis instead of SH basis. The HSH basis uses shifted associate Legendre polynomials to describe a 2D function projected onto a hemisphere $\mathbb{S}^2_{\theta_c \leq \pi/2}$ with a free edge.
\par
Huang et al. (2006) \cite{REF:18} was the first to apply HSH functions to extract shape descriptors for anatomical structures with free edges via a modified parameterisation algorithm of the original one proposed by \cite{REF:1}. By applying the proposed method to images of ventricles, they concluded that it is more robust and natural than the SH basis for representing free edges, which requires more effort to satisfactorily converge along the discontinuities. In addition to having the naturally defined free edge in HSH, the convergence is also faster than with the ordinary SH because of the distortion introduced by the parameterisation algorithm and the maximum wavelength defined on the regionally-selected patch on the targeted object, which will be discussed and demonstrated in later sections. As a similar strategy, Giri et al. (2020) \cite{REF:19} recently developed a hemispherical area-preserving parameterisation algorithm and analysed $60$ hemispherical-like anatomical surfaces. They compared the results with ordinary SH and also found significant improvements from $50$ to $80\%$ on the reconstruction quality. They also proposed variants of HSH for other anatomical shapes~\cite{giri2020brain,choi2021adaptive}.

\subsection{Regional modelling of spatial frequencies on a sphere}
\par
This paper uses SCH basis functions to study the local roughness and morphological features of rough surface patches that are regionally distributed over, for example, natural rocks, stones, bones, faults, etc. The idea of applying an SCHA to a regional subset of data distributed over a sphere was previously exploited in geophysics to study physical quantities such as geomagnetic and gravitational fields (refer to \cite{REF:21} for a comprehensive review of other applications). Many other methods can also be found in the literature for dealing with regional data.
\par
Multitaper spectral analysis is one common method for localising data using global basis functions (integrated over the whole domain) such as the spherical harmonics. In this method, the data outside the region of interest is tapered to zero. When the region of interest is a patch with a free edge, it can be parameterised over an appropriate cap with an assumption of a grid-of-zeros over the rest of the sphere. This technique can be implemented efficiently using sparse matrices; however, the data tapering will affect the analysis results \cite{REF:20}, and special relationships will be needed to link the localised analysis to the global one \cite{REF:70}.
\par
Another common approach is wavelet analysis, where the signals are multiplied (inner product) with a wavelet function and then integrated over the entire domain. However, this analysis becomes a function of the wavelet size (scale) and not the degrees of the SH expansion \cite{REF:20}. Although it is tempting to use these approaches, and their computation is fast and stable, they alter the actual signals and properties of the basis functions. Instead, we herein consider a direct and analytical approach that defines global functions over the entire regional domain without altering the actual data or basis functions.

\subsection{Spherical cap harmonic (SCH) basis functions}
\label{SUBSEC:REVIEW_SCH}
\par
The SCH basis functions were first proposed by Haines (1985) \cite{REF:22}, and his codes with the implementation details were published shortly thereafter~\cite{REF:24}. In his paper, he solved the Sturm–Liouville (self-adjoint) boundary value problem to derive two mutually orthogonal sets over a constrained spherical cap that satisfy the Laplacian. By applying one of the Neumann or Dirichlet boundary conditions on the free edge of the cap, we can get a general form of the associated Legendre function of real fractional (non-integer) degrees and integer orders. To satisfy the applied boundary conditions, we first solve for eigenvalues of the Sturm–Liouville boundary value problem, and these real and non-integer degrees of the associated Legendre functions can be used to define the basis functions. The resulting basis set depends on the applied boundary conditions; two types of bases are widely used and are named the \say{even} and the \say{odd} basis functions (for more details, see Section \ref{SEC:SCH}).
\par
Haines's basis functions have been studied and revised several times. In general, they were derived to analyse signals on a solid spherical cap with a finite thickness, which is described by an internal ($r = a$) and external ($r = b$) radii. These basis functions were used to expand physical quantities through the Lithosphere. However, in this work, we use the basis for a spherical cap with zero thickness, i.e. $a \to b$. This simplifies the problem because the normalisation of the spherical cone is ignored and the associated extra boundary conditions do not need to be applied through the thickness $b-a$. Haines's derivation of the normalisation factor was indeed an approximation (Schmidt-normalised), and he claims that the normalisation itself is not crucial as soon as the same normalisation factor is used for the reconstruction. An analytical derivation for the normalisation factor was proposed by Hwang et al. (1997) \cite{REF:23}, and they named it the fully normalised SCH.
\par
To capture smaller minimum wavelengths at the surface of the earth than with the original SCH, De Santis (1991) \cite{REF:26} proposed a modified version developed by shifting the origin of the spherical cap from the origin of the earth towards the surface. In 1992, De Santis \cite{REF:25} proposed using the ordinary SH defined over a hemisphere (using integer degrees) by scaling the spherical cap data to fit over a hemisphere. Although it is tempting to use such approximations to relax the complexity of the proposed functions in the original SCH, they are only valid for small half-angles (shallow spherical caps). More recently, Th\'ebault and his co-authors \cite{REF:27, REF:28} proposed a new revised version of the SCH to deal with problems mainly associated with the solid spherical cone. These papers are considered out-of-scope here because we only deal with the surfaces of spherical caps. More details about the history and the evolution of the method and its applications in geophysics can be found in \cite{REF:21, REF:29}.

\subsection{Parameterisation algorithms}
\par
Before expanding spatial data using the SCH basis, the surface patch needs to be mapped onto the spherical cap. This mapping algorithm assigns each vertex coordinate of the surface of the original object a unique coordinate on the spherical cap. A proper parameterisation algorithm should distribute the morphological features uniformly over the surface without clustering \cite{REF:1}. The mapping algorithms used here can be categorised into angle-preserving (conformal)~\cite{jin2004optimal,choi2018linear} or area-preserving~\cite{zhao2013area,choi2018density}. The former preserves the angles between the mesh elements such that that the local morphological properties are preserved, but this could introduce undesirable distortion in the element areas. The latter algorithm preserves the area of the elements but can introduce sufficient angular distortion to jeopardise the representation of the local morphological features. In general, angle and area preservation cannot be achieved at the same time~\cite{floater2005surface}, hence the choice of parameterisation method depends on the application. In mesh fitting, it is often desirable to balance angle and area distortions \cite{REF:30}.

\section{Spherical cap harmonic (SCH) basis functions}
\label{SEC:SCH}
In this section, we describe the set of orthogonal functions on the surface of the  spherical cap that we use for studying the geometry of a nominally flat and open-edged patch projected onto the cap. To do this, we propose using the basis functions of the SCHs that satisfy the Laplace equation in 3D; the derivations for these functions can be found in~\cite{REF:22, REF:23}. Here, we introduce the derivation only to the extent that the definition and meaning of the various components of the functions are clear. It serves also as documentation for the accompanying code, which is shared openly and in which these equations have been implemented numerically.
\par
Here, we are considering only polar spherical caps, meaning the local polar axis of the cap coincides with the global polar axis of the unit sphere. The coordinates of any point on the unit spherical cap are defined in terms of the angles $\theta$ and $\phi$. The angle $\theta$ represents the latitude (elevation angle), measured from the positive z-axis $\in [0, \theta_c]$, and the angle $\phi$ represents the azimuth angle (longitude), measured counter-clockwise from the positive x-axis $\in [0, 2\pi]$. The size of the spherical cap $\mathbb{S}^2_{~\theta \leq \theta_c}$ is therefore characterised by the half-angle of the spherical cap $\theta_c$, which is the latitude of the rim of the spherical cap as illustrated in Figure \ref{FIG:spherical_cap}.
\par
A 2D signal $f(\theta, \phi)$, for which each value is paired with $(\theta, \phi)$, can be written using the SCH basis as (see \cite{REF:22}):
\begin{equation}
    f(\theta, \phi) = \sum_{k = 0}^{\infty} \sum_{m = -k}^{k} {}^{\theta_c} q^m_k~ {}^{\theta_c} C_{k}^{m} (\theta, \phi),
\end{equation}
where $^{\theta_{c}} C_{k}^{m} (\theta, \phi)$ are the SCHs and ${}^{\theta_c} q^m_k$ are specific constants (SCH coefficients). The complex-valued SCHs of degree $l$ and order $m$ for a cap of size $\theta_c$ can be expressed as:
\begin{equation}
   ^{\theta_{c}} C_{k}^{m} (\theta, \phi) = \underbrace{\Bar{P}_{l(m)_k}^{m} (cos\theta)}_{\text{ALF}} \overbrace{~e^{im\phi}~}^{\text{Fourier}}.
\end{equation}
One way of solving the Laplace equation is by separating the independent variables, $\theta$ and $\phi$ in our problem, into separate ordinary differential equations and then assuming the overall solution to be the product of the separated solutions. The SCHs can therefore be written as a product of the term that describes the variation with regard to the latitude $\theta$ and a term that describes the variation with regard to the longitude $\phi$. In polar caps, the longitudinal direction ($\phi$) of the spherical cap covers, by definition, complete circles (\textit{Parallels of Latitude}), so we apply the traditional Fourier method to capture the circular variations.
\par
The variations along the latitudes, represented by ($\theta$), are captured using, for example, associated Legendre functions (ALF) as a mutually orthogonal set of basis functions. In the following subsections, we describe the ALFs, how to stably compute them and how the degrees $l$s of these ALFs are found. Then we explain one possible normalisation factor that we used in this paper.

\subsection{Associated Legendre function (ALF)}
\label{SUBSEC:ALF}
\par
In classical SH, the degree $l$ and the order $m$ are integers, and the associated Legendre polynomials (ALPs) are defined for the interval $x \in [-1, 1]$. In this interval, they form an orthogonal set of equations on a unit sphere (for $x = \cos\theta, \theta \in [0, \pi]$). In the HSH basis functions, a linear transformation shifts the orthogonality interval to either $[0, 1]$ or $[-1, 0]$ depending on whether the lower $\mathbb{S}^2_{~\theta \geq \pi/2}$ or upper $\mathbb{S}^2_{~\theta \leq \pi/2}$ hemisphere is considered (see the works in \cite{REF:18, REF:19}). The hemisphere introduces a new boundary (free edge) at the equator that must be defined with an additional appropriate boundary condition. However, note that the tangent at the edge of the equator (the gradient with respect to $\theta$) is zero. As will be seen next, this is one of the Neumann boundary conditions for the boundary value problem of ALFs, which makes it a special case of the SCH basis with integer degrees and orders. The main challenges with SCHs are shifting the ALFs to be defined over the interval $[\cos \theta_c, 1]$ and assuring their orthogonality.
\par
Using SH or HSH functions to describe a spherical cap (sub-region over the whole valid domain) will produce a set of ill-conditioned equations \cite{REF:23}. Although these sets could be used to some extent to reconstruct the input surfaces, the power spectral analysis will be chaotic and without the clear attenuating trends normally produced when the equations are well-conditioned. To find a suitable set of basis functions, Haines \cite{REF:22} proposed treating the problem as an inverse problem and determining the degrees $l$ for the ALPs that produce an orthogonal basis over the spherical cap while satisfying the boundary conditions. These degrees are not constrained to integers, and in our case, the degrees will be real-valued numbers. The orders $m$ are still integers because, for polar caps, $\phi$ always varies over complete circles.
\par
The basis functions proposed by Haines (1985)~\cite{REF:22} satisfy three boundary conditions: (i) continuity along the longitudes, (ii) regularity at the pole of the cap, and (iii) an arbitrary signal value at the free-edge (see \cite{REF:39, REF:21}). The first boundary condition is assured by the use of polar spherical caps (notice the complete circles in Figure \ref{FIG:spherical_cap}), as they force the periodicity and the continuity along the longitudes to make the order $m$ a positive or negative integer:
\begin{equation}
    f(\theta, \phi) = f(\theta, \phi + 2\pi),
    \label{EQN:BC1_1}
\end{equation}
\begin{equation}
    \frac{\partial f(\theta, \phi)}{\partial \phi} = \frac{\partial f(\theta, \phi + 2\pi)}{\partial \phi}.
    \label{EQN:BC1_2}
\end{equation}
\par
The second boundary condition, the regularity at the cap's pole $\theta=0$, is satisfied through the use of the first kind of ALFs, such that:
\begin{equation}
    f(\theta = 0, \phi) = 0 \ \ \  \text{ if } m \neq 0,
    \label{EQN:BC2_1}
\end{equation}
\begin{equation}
    \frac{\partial f(\theta = 0, \phi)}{\partial \phi} = 0 \ \ \text{ if } m = 0.
    \label{EQN:BC2_2}
\end{equation}
\par
{Note that even with the two above boundary conditions, the solution functions are still not unique. To obtain unique solution functions, we need to impose a proper boundary condition at the edge of the spherical cap.} The third boundary condition is applied to the free edge at $\theta_c$ ($x_c = \cos \theta_c$):
\begin{equation}
    A_c P^m_{l(m)_k}(x_c) + B_c \frac{d P^m_{l(m)_k}(x_c)}{dx} = 0.
    \label{EQN:BC}
\end{equation}
Here, we use the degree notation $l(m)_k$ instead of $l$ because the degrees are now back calculated for a specific order $m$ and index $k$, which will be explained in detail below.
\par
In Eq.~(\ref{EQN:BC}), the case of $A_c = 0$ and $B_c \neq 0$ corresponds to the application of the Neumann boundary condition where the gradient with respect to $\theta$ at the cap's rim at $x = \cos \theta$ is zero. It implies that $f(\theta_c, \phi) \neq 0~ \forall \phi \in [0, 2\pi]$ {and can take any arbitrary value at the edge of the spherical cap}. Applying this boundary condition results in basis functions referred to by Haines (1985) \cite{REF:22} as the \say{even} ($k-m$ = even) basis or by Hwang et al. (1997)~\cite{REF:23} as set $2$. The case of $A_c \neq 0$ and $B_c = 0$ corresponds to the Dirichlet boundary conditions $f(\theta_c, \phi) = 0 \ \ \forall \phi \in [0 , 2\pi]$, and the resulting basis functions are known as the \say{odd} ($k-m$ = odd) set or set $1$ by Haines \cite{REF:22} or Hwang et al. \cite{REF:23}, respectively. The boundary conditions {for the two cases} can therefore be rewritten as:
\begin{equation}
    \frac{d P^m_{l(m)_k}(x_c)}{dx} = 0~~\text{for $k-m=$ even},
    \label{EQN:BC2_a}
\end{equation}
\begin{equation}
    P^m_{l(m)_k}(x_c) = 0~~\text{for $k-m=$ odd}.
    \label{EQN:BC2_b}
\end{equation}
\par
The latter boundary condition{, Eq.~(\ref{EQN:BC2_a}) or (\ref{EQN:BC2_b}),} is met by back calculating the roots $l(m)_k$ at different orders $m$ and indices $k$ {and then using them for} constructing the basis functions for different $\theta_c$. Although it is tempting to use a combination of both sets to expand a signal (e.g. see Haines~\cite{REF:22}), for the uniform convergence of the series, it is noteworthy that these basis functions are mutually orthogonal and cannot be used together for the power spectral analysis \cite{REF:44}. In this paper, we will only consider the even set, as it allows arbitrary values for the signal at the cap's rim ($f(\theta_c, \phi) \neq 0~ \forall \phi \in [0, 2\pi]$), which contrasts with the odd set that equates the signal at the rim to zero. This consideration therefore allows us to correctly reconstruct any surface geometry.
\par
Unlike the traditional ALPs, here the degrees $l(m)_k$ are determined such that the third boundary condition is satisfied; thus, $l(m)_k$ are real numbers, though not necessarily integers; hence, we use the index $k$ to rank the orders. $k$ is an integer index that arranges $m$ from $0 \text{ to } k$, and the degree $l(m)_k$ is the degree at the $k^{\text{th}}$ index and order $m$. The $k$ index is useful for finding the degrees on the applied boundary conditions. For instance, when $k=0$, we find the first degree $l(m)_k \geq 0$ where the boundary condition equation first crosses zero (first eigenvalue).
\par
For constructing the ALFs with integer orders and real degrees, Hobson (1965) \cite{REF:38} proposed the use of hypergeometric functions. Note that in the following, we refer to associated Legendre \textit{functions} rather than \textit{polynomials} because these basis functions are truncated numerically at a predefined numerical threshold. These functions can also be referred to as fractional ALFs, because the associated degrees are not integers. The ALFs can then be written as follows (see~\cite{REF:22, REF:23, REF:38}):
\begin{equation}
    P^m_l (x) = \frac{1}{2^m m!} \frac{\Gamma(l+m+1)}{\Gamma(l-m+1)} (1-x^2)^{\frac{m}{2}}
    _2F_1\Big(m-l, m+l+1; m+1; \frac{1-x}{2}\Big),
\label{EQN:ALF}
\end{equation}
where $\Gamma(\cdot)$ is the gamma function, $_2F_1(a, b; c; z)$ is the hypergeometric function defined over a unit disk $|Z|<1$ and $x = \cos\theta$.
\par
To find the zeros (eigenvalues) that satisfy the third boundary condition, we rewrite Legendre's differential equation in a Sturm-Liouville (S-L) form \cite{REF:22, REF:23}:
\begin{equation}
    \frac{d}{dx}\left[(1-x^2) \frac{dP^m_l(x)}{dx}\right] + \left[ l(l+1) - \frac{m^2}{1-x^2}\right]  P^m_l(x) = 0
    \label{EQN:SL_PDE}
\end{equation}
and subject it to the boundary conditions explained in Eq.~(\ref{EQN:BC}). The two equations that result from the application of the Neumann and Dirichlet boundary conditions can then be written in a simpler factorised form, as proposed by Hwang et al. (1997)~\cite{REF:23} for hypergeometric functions:
\begin{equation}
    l(m)_k x_c F\big(l(m)_k, m, x_c\big) - \big(l(m)_k-m\big) 
    F\big(l(m)_k - 1, m, x_c\big) = 0,
    \label{EQN:BC_1}
\end{equation}
\begin{equation}
    F\big(l(m)_k, m, x_c\big) = 0,
    \label{EQN:BC_2}
\end{equation}
where
\begin{equation}
    F(l, m, x) = _2F_1 \Big(m-l, m+l+1; m+1; \frac{1-x}{2}\Big).
\end{equation}
Equation~(\ref{EQN:BC_1}) is the equivalent of applying the Neumann boundary conditions (even set), and Eq.~(\ref{EQN:BC_2}) corresponds to the Dirichlet boundary conditions (odd set).
\par
Both Eqs.~(\ref{EQN:BC_1}) and (\ref{EQN:BC_2}) have an infinite number of roots (eigenvalues) for the above-mentioned S-L problem. Here, we are interested in finding the first $k+1$ eigenvalues $l(m)_k$ where $k\in \{0, \dots, |m|\}$, which is done using a simple stepping solver based on Mueller's method for finding zeros (roots) \cite{REF:66}. Additional results about the roots for $\theta_c = 5\pi/18$ are shown in Appendix \ref{SUBSEC:SL_eign}.
\par
\begin{figure}
\centering
\includegraphics[width=0.65\textwidth]{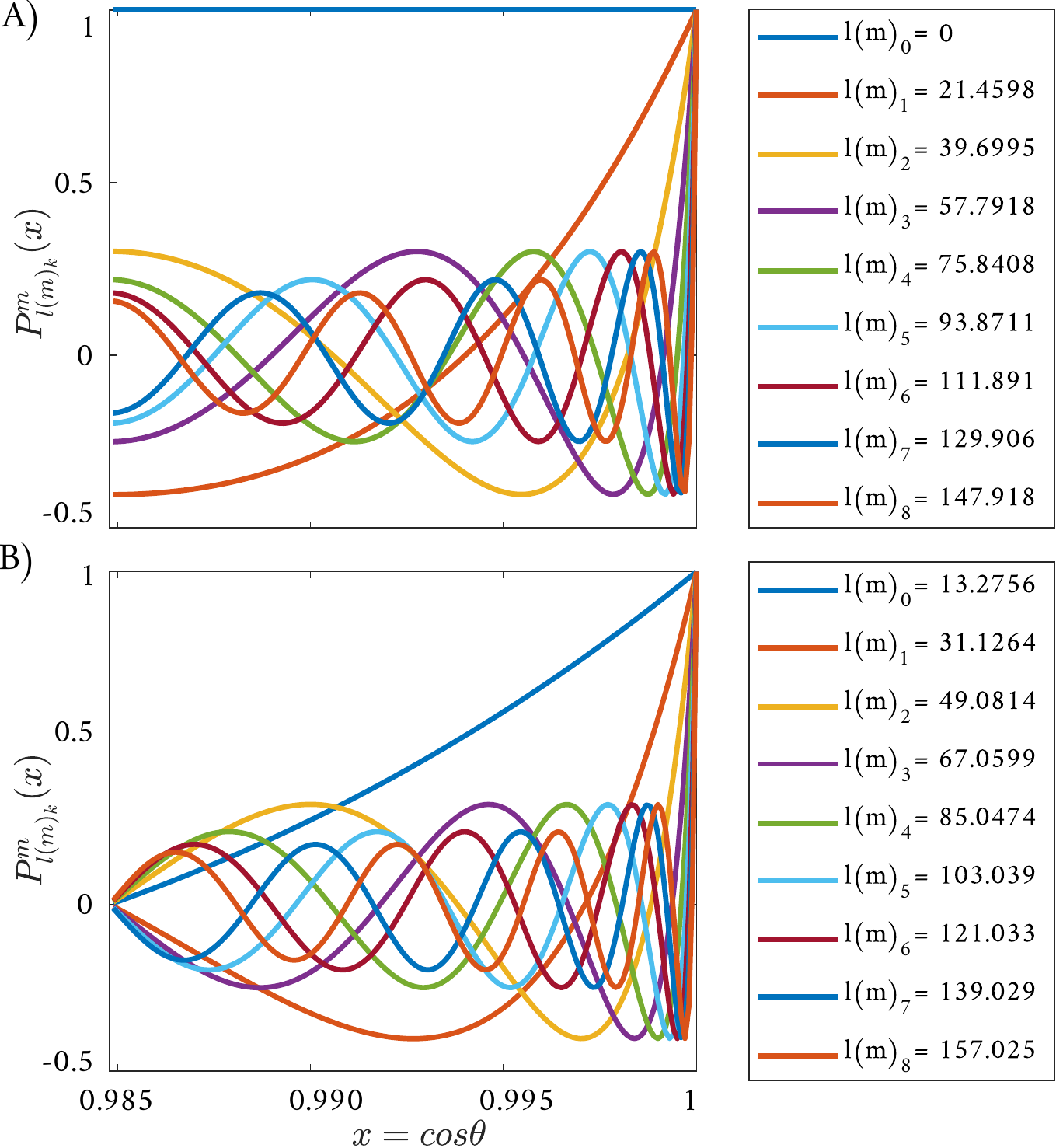}%
\caption{The associated Legendre functions (ALFs) for $\theta_c = \pi/18$, $m = 0$ and $k \in \{0, 1, \dots, 8\}$; (A) the even basis and the corresponding eigenvalues from Eq.~(\ref{EQN:BC_1}); (B) the odd basis and the corresponding eigenvalues from Eq.~(\ref{EQN:BC_2}).
\label{FIG:legendre_basis}
}
\end{figure}

After finding the needed roots for the different $m$ and $k$, we can substitute these roots directly into Eq.~(\ref{EQN:ALF}) to obtain the needed ALF basis. Figure \ref{FIG:legendre_basis} shows the ALFs associated with the roots for different $k$, where $m = 0$ and $\theta_c = 10^{\circ} = \pi/18$ (i.e. $\cos(\pi/18) = 0.985$). Graphically, the value $k-m$ indicates how many times the basis function changes signs along the interval $[\cos\theta_c, 1]$; the $(k+1)^{\text{th}}$ eigenvalue changes signs once more than the $k^{\text{th}}$ root, as it divides the interval into regions with different signs. Together, $k \text{ and } m$ define the frequency classification (harmonics) of the basis functions as zonal ($m = 0$), sectoral ($|m| = k$) or tesseral ($0<|m|<k$) regions on the SCH, as will be illustrated next. Figure \ref{FIG:legendre_basis} also shows the difference between even and odd sets depending on the applied boundaries at $x = \cos \theta_c$.
\par
Equations~(\ref{EQN:ALF}), (\ref{EQN:BC_1}) and (\ref{EQN:BC_2}) are dependent on the ordinary Gaussian hypergeometric function evaluation $_2F_1(a,b;c;z)$. Unfortunately, the implementations given in \cite{REF:22, REF:23} using the recursive series are not numerically stable for large degrees when $k>12$ about the radius of convergence $|Z|<1$. Indeed this is a well-known problem for such functions, and different formulations can be found in the literature with different implementations for different ranges of $a$, $b$, $c$ and $z$ (see Pearson et al. [2015] \cite{REF:67}, who reviewed different hypergeometric functions and their different implementations and stability conditions). Therefore, we herein use a Taylor series expansion for the first degrees (up to $k = 12$) when $a$, $b$ and $c$ are relatively small. Otherwise, we use a $5^{\text{th}}$ order Runge-Kutta algorithm along with the Dormand-Prince method for solving the hypergeometric differential equation (see \cite{REF:67} for more details).

\subsection{Normalising the SCH basis functions}
\par
Normalising the basis functions is important for ensuring their orthonormality and for keeping the calculation of the coefficients computationally manageable \cite{REF:23, REF:39}. Many normalisation methods have been proposed in the literature, such as the \say{Heiskanen and Moritz} method \cite{REF:23} and the \say{Schmidt} or \say{Neumann} methods \cite{REF:38}, and any of these can be used as long as the same normalisation method is used for both the expansion and the reconstruction \cite{REF:22}. In the following, we use the \say{Schmidt semi-normalised} method by Haines \cite{REF:22}. 
\par
This normalisation method requires solving the integral for the ALFs over the spherical cap (mean square product), which can be retrieved from \cite{REF:41} with a corrected sign from \cite{REF:40}:
\begin{equation}
    \int_{0}^{\theta_c} \Big[P^m_{l}(\cos\theta)\Big]^2 \sin\theta\, dx 
    = \Bigg| \sin\theta \Bigg( \underbrace{\frac{\partial P^m_{l}}{\partial l}\frac{\partial P^m_{l}}{\partial \theta}}_{\text{odd}}  -  \underbrace{P^m_{l} \frac{\partial }{\partial l} \frac{d P^m_{l}}{d \theta}}_{\text{even}}\Bigg) \Bigg|_0^{\theta_c} \frac{1}{2n +1}.
\end{equation}
However, this complex expression is difficult to solve (see \cite{REF:23} for one approach using recursive expressions). In this paper, we use the Schmidt semi-normalised harmonics introduced by Haines (1985)~\cite{REF:22}:
\begin{equation} \label{EQN:NORM_FACTOR}
K_{l(m)_k}^m =  \left\{
\begin{array}{ll}
      1, & ~~m = 0 \\
      \frac{k 2^{-m}}{\sqrt{m\pi}} \Big( \frac{l(m)_k + m}{l(m)_k - m} \Big)^{\frac{l(m)_k}{2} + \frac{1}{4}},& ~~m > 0 \\
\end{array}
\right.
\end{equation}
where
\begin{equation}
    k = p^{\frac{m}{2}} e^{e1 + e2 + \dots},
\end{equation}
with
\begin{equation}
    p = \Bigg( \frac{l(m)_k}{m}\Bigg)^2,
\end{equation}
\begin{equation}
    e1 = -\frac{1}{12m}\Big( 1 + \frac{1}{p} \Big), \\
\end{equation}
\begin{equation}
    e2 = \frac{1}{360m^3}\Big( 1 + \frac{3}{p^2} + \frac{4}{p^3}\Big).
\end{equation}
\par
The final form of the normalised ALFs can be written as follows (see~\cite{REF:22}):
\begin{equation}
    \Bar{P}^{m}_{l(m)_k} (x) = K_{l(m)_k}^m (1-x^2)^{\frac{m}{2}}
    _2F_1\Big(m-l, m+l+1; m+1; \frac{1-x}{2}\Big).
    \label{EQN:NORM_ALF}
\end{equation}
We use Eq.~(\ref{EQN:NORM_ALF}) as a part of the final SCH basis functions.

\subsection{The SCH series}
\par
The complete form of the SCH can then be expressed by combining Legendre's functions with Fourier expansions:
\begin{equation}
   ^{\theta_{c}} C_{k}^{m} (\theta, \phi) = \underbrace{\Bar{P}_{l(m)_k}^{m} (\cos\theta)}_{ALF} \overbrace{~e^{im\phi}~}^{Fourier}.
   \label{EQN:SCF1}
\end{equation}
We can split this expression into a real part and an imaginary part:
\begin{equation}
\left\{
\begin{array}{ll}
      \mathfrak{R}\Big( {}^{\theta_{c}} C_{k}^{m} (\theta, \phi) \Big) = \Bar{P}_{l(m)_k}^{m} (\cos\theta) \cos(m\phi),\\~\\
      \mathfrak{I}\Big( {}^{\theta_{c}} C_{k}^{m} (\theta, \phi) \Big) = \Bar{P}_{l(m)_k}^{m} (\cos\theta) \sin(m\phi).\\
\end{array} 
\right. 
\end{equation}
\par
\begin{figure*}
\centering
\includegraphics[width=1.0\textwidth]{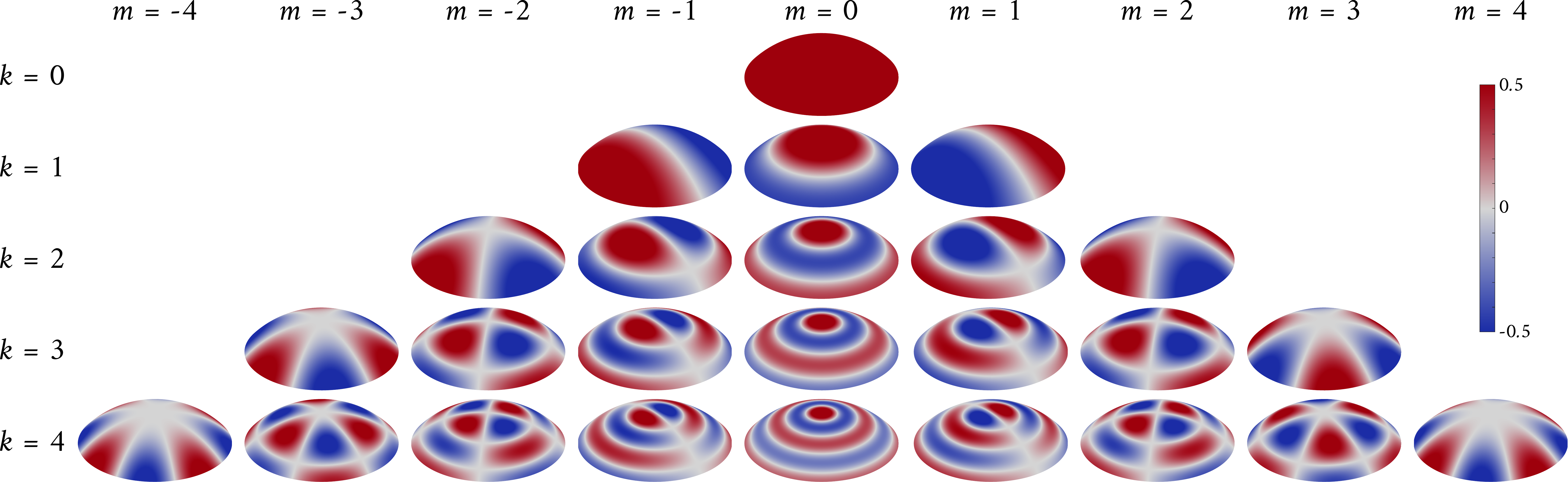}%
\caption{The normalised spherical cap harmonic (SCH) basis functions up to $k = 4$ at $\theta_c = 5\pi/18$. The figures show $\mathfrak{R}\Big(^{5\pi/18} C_{k}^{m} (\theta, \phi) \Big)$. Note that at the breathing mode when $k = m = 0$, the surface inflation is unity and is therefore not included in the colorbar. The complex part of the basis are similar but rotated $\pi/(2|m|)$ about the $z$-axis.
\label{FIG:SCH_basis_map}
}
\end{figure*}

The complex-valued equation (\ref{EQN:SCF1}) is only valid for non-negative orders $|m|$. To address this, the redundant negative-orders equation can be written in terms of the positive ones, as they reassemble $\pi/2$ rotations about the positive z-axis. By making use of the Condon-Shortly phase factor, we can write the SCH functions for $m<0$ as:
\begin{equation}
   ^{\theta_{c}} C_{k}^{-m} (\theta, \phi) = (-1)^m~ ^{\theta_{c}} C_{k}^{m *} (\theta, \phi),
   \label{EQN:SCF2}
\end{equation}
where $C_{k}^{m *} (\theta, \phi)$ is the complex conjugate of the SCH functions with positive orders. Figure \ref{FIG:SCH_basis_map} shows the real SCH functions (with positive and negative orders) for the half-angle $\theta_c = 5\pi/18$ and up to $k = 4$.

\par
We can write a bi-directional signal $f(\theta, \phi)$ that is explicitly parameterised to unique $\theta \in [0, \theta_c]$ and $\phi \in [0, 2\pi]$ over a unit spherical cap $\mathbb{S}^2_{~\theta \leq \theta_c}$ as a linear sum of the independent harmonics defined over the spherical cap:
\begin{equation}
    f(\theta, \phi) = \sum_{k = 0}^{\infty} \sum_{m = -k}^{k} {}^{\theta_c} q^m_k~ {}^{\theta_c} C_{k}^{m} (\theta, \phi),
    \label{EQN:SERIES1}
\end{equation}
where ${}^{\theta_c} q^m_k$ are the complex-valued SCH coefficients at order $m$ and index $k$. In our case, following Brechb\"uhler's work in \cite{REF:1}, the signal $f(\theta, \phi)$ is the 3D vector that comprises three orthogonal components in the Cartesian space extracted from the coordinates of the vertices:
\begin{equation}
    f(\theta, \phi) = 
        \begin{pmatrix}
        x(\theta, \phi)\\
        y(\theta, \phi)\\
        z(\theta, \phi)
        \end{pmatrix}.
    \label{EQN:CART_VECTOR}
\end{equation}
In practice, we truncate the series in Eq.~(\ref{EQN:SERIES1}) at $k = K_{\max}$, which reassembles the minimum-accumulated wavelength (equivalent to the cut-off frequency) $\omega_{\min}$ in the harmonic sum from the original surface patch. Then, $f(\theta, \phi)$ can be approximated by:
\begin{equation}
    f(\theta, \phi) \approx \sum_{k = 0}^{K_{\max}} \sum_{m = -k}^{k} {}^{\theta_c} q^m_k~ {}^{\theta_c} C_{k}^{m} (\theta, \phi).
    \label{EQN:SERIES2}
\end{equation}
The coefficients ${}^{\theta_c}q^m_k$ can be expanded as:
\begin{equation}
    {}^{\theta_c}q^m_k =~ <f(\theta, \phi)~ {}^{\theta_c}C_{k}^{m} (\theta, \phi)>,
    \label{EQN:COEF_INTG1}
\end{equation}
\begin{equation}
    {}^{\theta_c}q^m_k = \int_0^{\theta_c} \int_0^{2\pi} f(\theta, \phi)~ {}^{\theta_c}C_{k}^{m} (\theta, \phi) \sin\theta \,d\phi \,d\theta.
    \label{EQN:COEF_INTG2}
\end{equation}
However, this integral is valid only when the signal $f(\theta, \phi)$ is defined everywhere over the spherical cap (continuous signal). For discrete signals, we can estimate the coefficients using least-square fitting algorithms, as will be discussed later in Section \ref{SEC:SCHA}.

\section{Spherical cap conformal parameterisation}
\label{SEC:PARAMETRIZATION}
Here we develop a conformal parameterisation method for mapping any surface $\mathcal{S}$ with disk topology to the spherical cap $\mathbb{S}^2_{~\theta \leq \theta_c}$, where $\theta_c$ is the prescribed half angle. The overall spherical cap parameterisation $f:\mathcal{S} \to \mathbb{S}^2_{~\theta \leq \theta_c}$ is desired to be not only conformal but also with low area distortion.

First, note that the south-pole stereographic projection $\tau$ given by
\begin{equation} \label{eqt:south_stereographic}
    \tau(x,y,z) = \left(\frac{x}{1+z}, \frac{y}{1+z}\right)
\end{equation}
maps the spherical cap $\mathbb{S}^2_{~\theta \leq \theta_c}$ to a planar disk $\mathcal{D}_r$ centred at the origin with radius 
\begin{equation}
    r = \sqrt{\frac{1 - \cos \theta_c}{1+\cos \theta_c}},
\end{equation}
and the inverse south-pole stereographic projection $\tau^{-1}$ given by
\begin{equation} \label{eqt:inverse_south_stereographic}
        \tau^{-1}(X,Y) \\
        = \left(\frac{2X}{1+X^2+Y^2}, \frac{2Y}{1+X^2+Y^2},\frac{1-X^2-Y^2}{1+X^2+Y^2}\right)
\end{equation} 
maps $\mathcal{D}_r$ to $\mathbb{S}^2_{~\theta \leq \theta_c}$. As both $\tau$ and $\tau^{-1}$ are conformal, we were motivated to find an optimal conformal mapping from the given surface $\mathcal{S}$ to the planar disk $\mathcal{D}_r$ so that we could apply the inverse projection $\tau^{-1}$ to obtain the desired parameterisation.

\begin{table*}[!ht]
\centering
\begin{tabular}{lccc}
\hline
PSS & $X$ & $Y$ & $d_{area}$ \\
\hline
mean   & 0.031115190925885 & 2.283277253594884 & 1.621661326243250 \\
median & 0.032048117800906 & 2.418244833527807 & 1.621524158857667 \\
std    & 0.005932670562184 & 0.753260686325562 & 3.56905230744E-04 \\
\hline
\end{tabular}
\caption{The results summary of $51$ consecutive runs of the PSS algorithm for the benchmarking test shown in Figure \ref{FIG:monkeyhead_benchmark}.}
\label{TAB:monkeyhead_benchmark}
\end{table*}

\begin{figure*}
\centering
\includegraphics[width=1.0\textwidth]{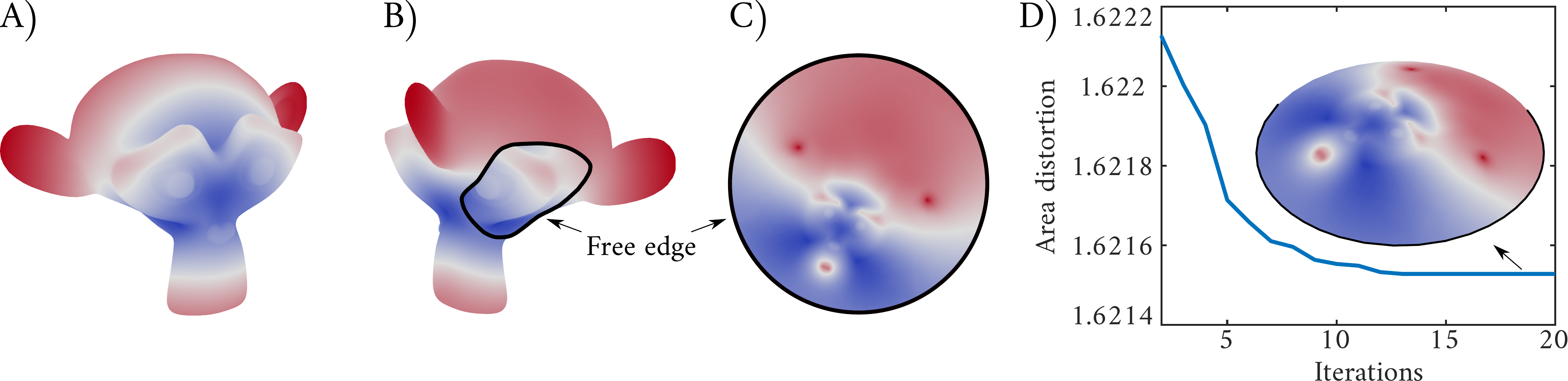}%
\caption{Parameterisation of the monkey head model. A) The monkey head (Suzanne) benchmark mesh~\cite{REF:37}, with the heat map (texture) illustrating the radial distance from the model's centroid; B) The back view of the model reveals the open boundary location; C) The conformal mapping onto the unit disk obtained using~\cite{REF:2}; D) The convergence curve for finding the optimal M\"obius transformation that minimises the area distortion on the spherical cap using the PSS algorithm $f:\mathcal{S} \to \mathbb{S}^2_{~\theta \leq \theta_c}$ for $\theta_c = 40^{\circ}$ (i.e. $r = 0.3640$).
\label{FIG:monkeyhead_benchmark}
}
\end{figure*}
\par
To achieve this, we first apply the disk conformal mapping method~\cite{REF:2} to map $\mathcal{S}$ onto the unit disk $\mathbb{D}$ in an angle-preserving manner (see~\cite{REF:2} for the algorithmic details). Denote the conformal mapping by $g:\mathcal{S} \to \mathbb{D}$. Now, we search for a conformal mapping $h:\mathbb{D} \to \mathcal{D}_r$ that can further reduce the distortion in area of the parameterisation. {Mathematically, a M\"obius transformation is a function that maps a complex number $z = X+Y i$ (where $X$, $Y$ are real numbers and $i$ is the imaginary number with $i^2 = -1$) to another complex number $(az+b)/(cz+d)$, where $a,b,c,d$ are complex coefficients with $ad-bc \neq 0$. M\"obius transformations are conformal and hence are a good candidate for the mapping $h$ in our problem. More specifically, here} we consider a M\"obius transformation $h:\mathbb{D} \to \mathcal{D}_r$ in the form 
\begin{equation}
h(X,Y) = r\frac{(X+Yi)- (A+Bi)}{1 - (A-Bi)(X+Yi)},
\end{equation}
where $A, B$ are two real numbers to be determined. To find the optimal $h$, we minimise the following area distortion measure
\begin{equation}
\label{EQN:AREA_DIST}
d_{\text{area}} = \text{mean}_{T \in \mathcal{F}} \left|\log \frac{\frac{\text{Area}(\tau^{-1}\circ h \circ g(T))}{ \sum_{T' \in \mathcal{F}} \text{Area}(\tau^{-1}\circ h \circ g(T'))}}{\frac{\text{Area}(T)}{\sum_{T' \in \mathcal{F}} \text{Area}(T')}} \right|,
\end{equation}
where $\mathcal{F}$ is the set of all triangular faces of $\mathcal{S}$. The desired spherical cap parameterisation is then given by 
\begin{equation}
f = \tau^{-1} \circ h \circ g.
\end{equation}
In particular, since every step described above is conformal, the overall mapping $f$ is also conformal. 
\par
{To explain the formulation of the area distortion measure $d_{\text{area}}$ in Eq.~\eqref{EQN:AREA_DIST} more clearly, we first consider the mathematical definition of area distortion. For any two surfaces $\mathcal{M}$, $\mathcal{N}$ in the Euclidean space $\mathbb{R}^3$ with the same total surface area, a mapping $f:\mathcal{M} \to \mathcal{N}$ is said to be area-preserving if for every open set $U$ on $\mathcal{M}$, the surface area of $U$ is equal to the surface area of $f(U)$. More generally, the area distortion of a mapping $f$ can therefore be evaluated using the dimensionless quantity $\left|\log \frac{\text{Area}(f(U))}{\text{Area}(U)}\right|$, which equals 0 if and only if $\text{Area}(f(U)) = \text{Area}(U)$. In our problem, $M$ corresponds to the input triangulated surface $\mathcal{S}$ and hence it is natural to evaluate the area distortion by considering every triangular face $T$. In other words, by minimising Eq.~\eqref{EQN:AREA_DIST} we look for a M\"obius transformation $h$ such that the area change of every triangular face under the spherical cap conformal parameterisation is as small as possible. Moreover, note that the total surface area of the input surface $\mathcal{S}$ is not necessarily equal to that of the target spherical cap domain $\mathbb{S}^2_{~\theta \leq \theta_c}$. Therefore, we use the two summation terms in Eq.~\eqref{EQN:AREA_DIST} for normalising the total surface area of the input surface and that of the spherical cap in the area distortion measure.}
\par
\begin{figure}
\centering
\includegraphics[width=0.65\textwidth]{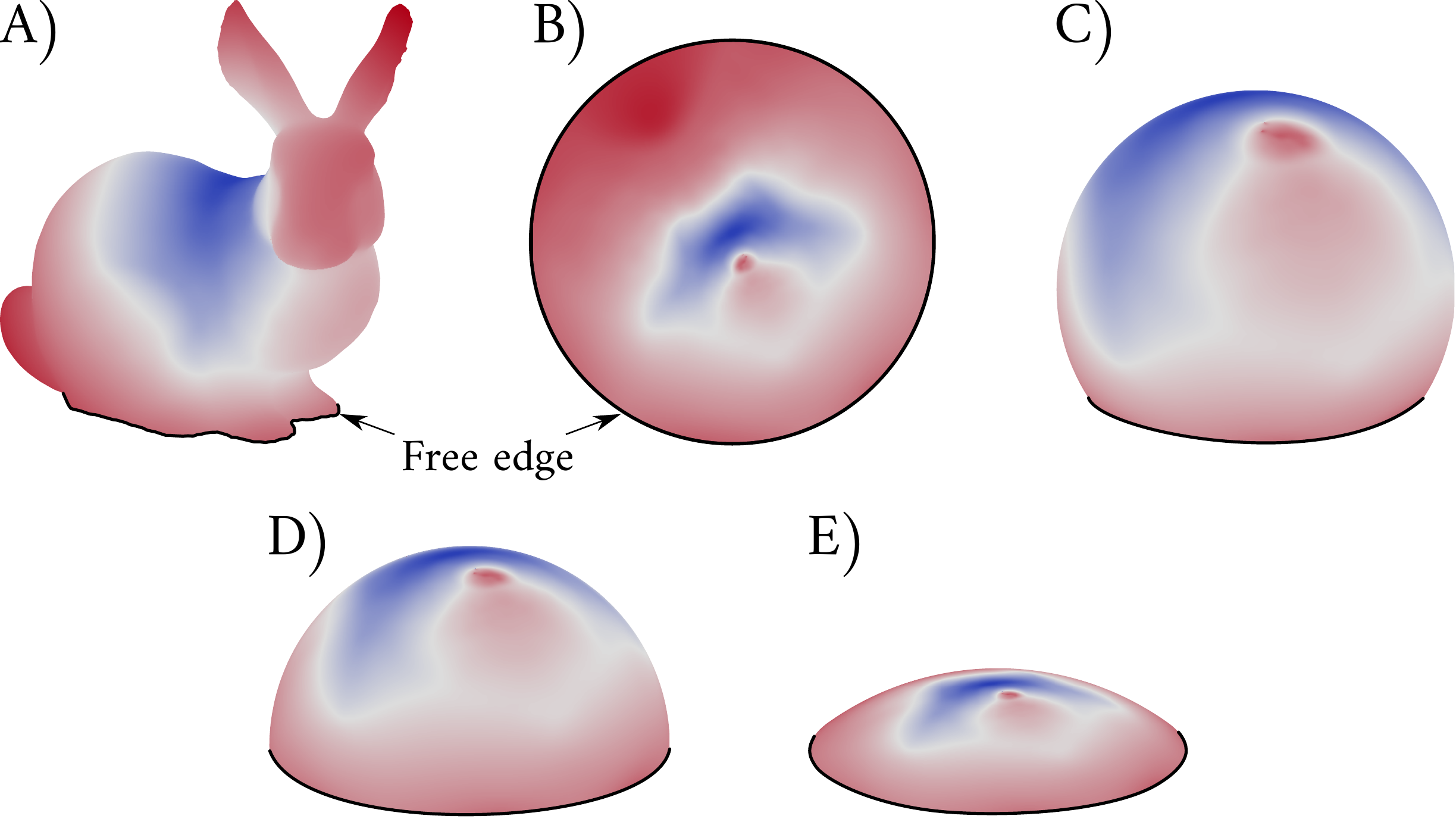}%
\caption{Spherical cap parameterisation of the Stanford bunny. A) The Stanford Bunny benchmark model~\cite{REF:Stanford_Bunny} with an open base, where the heat map (texture) illustrates the radial distance from the model's centroid; B) The conformal mapping onto the unit disk obtained using~\cite{REF:2}; C) The spherical cap parameterisation $f:\mathcal{S} \to \mathbb{S}^2_{~\theta \leq \theta_c}$ for $\theta_c = 120^{\circ}$ (i.e. $r = 1.7321$); D) The spherical cap parameterisation $f:\mathcal{S} \to \mathbb{S}^2_{~\theta \leq \theta_c}$ for $\theta_c = 90^{\circ}$ (a hemisphere, i.e. $r = 1.0$);
E) The spherical cap parameterisation $f:\mathcal{S} \to \mathbb{S}^2_{~\theta \leq \theta_c}$ for $\theta_c = 40^{\circ}$ (i.e. $r = 0.3640$).
\label{FIG:stanford_benchmark}
}
\end{figure}

In practice, the solution of this objective function can be found using a global search heuristic such as the Pareto-like Sequential Sampling (PSS) approach (see \cite{REF:65} for details). Using a population size of $30$ and an acceptance rate of $\alpha = 0.97$, only $20$ iterations are needed for most of the problems to reach a good approximation. This approach is gradient-free and does not require an initial value that could converge locally. Table \ref{TAB:monkeyhead_benchmark} shows the stability of the PSS approach, while Figure \ref{FIG:monkeyhead_benchmark} shows the convergence and the final result of the parameterisation step. Alternatively, the minimisation problem can also be solved using MATLAB's \texttt{fmincon}, using the modulus and argument of the complex number $A+Bi$ as variable. The solution obtained from this function was $d_{\text{area}} = 1.621523941729933$ with $X = 0.032055812166766$ and $Y = 2.416521766598053$, which is consistent with the PSS result.  {Figure~\ref{FIG:stanford_benchmark} shows several spherical cap parameterisation results with different prescribed $\theta_c$ for the Stanford bunny model~\cite{REF:Stanford_Bunny}. To further assess the conformality of the parameterisation results, we define the angle distortion $d_{angle}$ as the average of the absolute difference between every angle $[v_i,v_j,v_k]$ in a triangular face in the input surface and that in the spherical cap parameterisation $f$:
\begin{equation}
    d_{angle} = \underset{[v_i,v_j,v_k]}{\text{mean}} \frac{|\angle([f(v_i),f(v_j),f(v_k)]) - \angle([v_i,v_j,v_k])|}{180^{\circ}}.
\end{equation}
The values of $d_{angle}$ for the parameterisations in Figure~\ref{FIG:stanford_benchmark}(C), (D), and (E) are 0.00828069, 0.00869725, and 0.00919996 respectively. The results show that the proposed parameterisation method is highly conformal.}

\section{Numerical implementation of the spherical cap harmonic analysis (SCHA)}
\label{SEC:SCHA}
\par
In this section, we describe the numerical implementation of the SCHA method and use it to reconstruct triangulated surface meshes.

\subsection{Spherical cap harmonic analysis (SCHA)}
\par
SCHA comprises the computation of the weights ${}^{\theta_c}q^m_k$ by the integral in Eq.~(\ref{EQN:COEF_INTG2}). However, this integral cannot be calculated directly as the data is not defined everywhere over the spherical cap. Instead, it can be estimated for discrete points that are distributed randomly over the cap. For this reason, we estimate these coefficients using the least squares method (statistical regularisation). Following the indexing system proposed in \cite{REF:1}, the coefficients can be estimated as follows:
\begin{equation}
    {}^{\theta_c}q^m_k \approx (B^{T} B)^{-1} B^{T} V,
    \label{EQN:LSF}
\end{equation}
where $B$ contains the basis functions evaluated at the coordinates $V$ that define the geometry of the surface patch. If $n_{v}$ vertices describe the geometry of this surface patch in spatial coordinates, the matrix $V$ can be written as:
\begin{equation}
    V = \begin{pmatrix}
        x_0    & y_0    & z_0\\ 
        x_1    & y_1    & z_1\\ 
        \vdots & \ddots & \vdots\\
        x_{n_{v}-1} & y_{n_{v}-1} & z_{n_{v}-1}
    \end{pmatrix}.
    \label{EQN:VERTS}
\end{equation}
\par
It should be noticed that the vertices defined in Eq.~(\ref{EQN:VERTS}) are normalised against the mean, so the actual expanded signals will be identified about a mean of zero. The matrix $B$ holds the SCH basis functions that can be obtained by evaluating Eqs.~(\ref{EQN:SCF1}) and (\ref{EQN:SCF2}) for the positive and negative orders, respectively, and can be written as:
\begin{equation}
    B = \begin{pmatrix}
        c_{(0,0)} & c_{(0,1)} & \dots & c_{(0,j)}\\
        c_{(1,0)} & c_{(1,1)} & \dots & c_{(1,j)}\\
        \vdots & \vdots & \ddots & \vdots\\
        c_{(n_v-1,0)} & c_{(n_v-1,1)} & & c_{(n_v-1,j)}
    \end{pmatrix},
    \label{EQN:MAT_B}
\end{equation}
where $c_{i,j}$ is the basis function evaluated at the $i^{\text{th}}$ vertex for index $j$. The index $j$ is calculated as $j(m, k) := k^2 + k + m$ and $-k\leq m \leq k$ for the zero-based numbering programming convention. This means we need to fit $3(K_{\max}+1)^2$ coefficients simultaneously for all the $x(\theta, \phi)$, $y(\theta, \phi)$ and $z(\theta, \phi)$ signals in $f(\theta, \phi)$. The final complex-valued ${}^{\theta_c}q^m_k$ coefficient matrix can be written as:
\begin{equation}
    {}^{\theta_c}q^m_k = 
    \begin{pmatrix}
        q^0_{x}    & q^0_{y}    & q^0_{z}\\ 
        q^1_{x}    & q^1_{y}    & q^1_{z}\\ 
        \vdots & \ddots & \vdots\\
        q^j_{x} & q^j_{y} & q^j_{z}
    \end{pmatrix}.
    \label{EQN:Q_COEF}
\end{equation}
\par
The least squares fitting is a statistical regularisation method that provides a good approximations for the coefficients ${}^{\theta_c}q^m_k$, as long as the number of equations on the spherical cap is significantly larger than the expanded coefficients. Here, this condition is always satisfied because our surfaces are defined by dense point clouds. Other regularisation methods that enhance the accuracy of the least squares fitting and obtain a better orthogonality among the columns in Eq.~(\ref{EQN:MAT_B}) can be found in the literature (e.g. see \cite{REF:42}) but are not applied here.

\subsection{Spherical cap harmonic reconstruction}
\par
\begin{figure}
\centering
\includegraphics[width=0.7\textwidth]{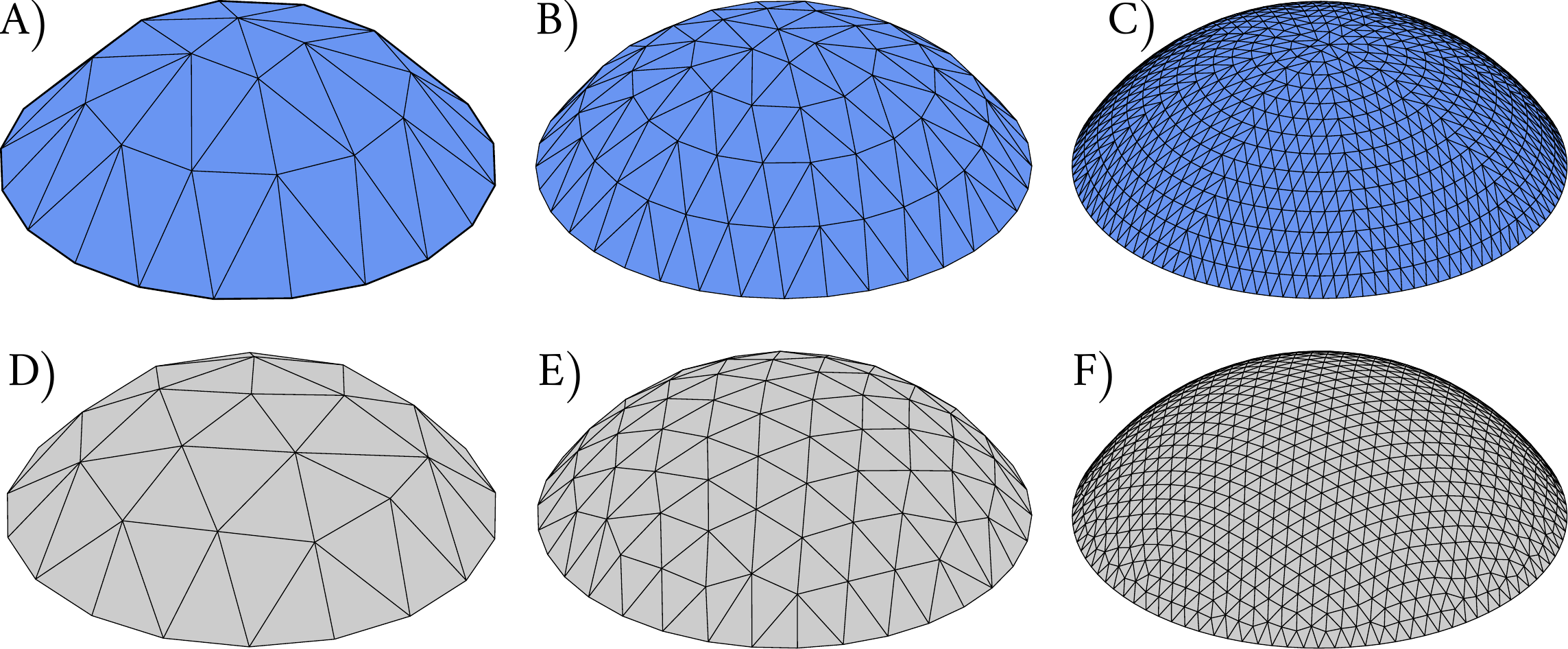}%
\caption{Geodesic domes for $\theta_c = \pi/3$ with different refinement cycles ($N$). The domes in A--C were obtained using the unit disk mesh generation method in \cite{REF:45} followed by the rescaling in Equation (\ref{EQN:SCALE_LAMBERT}) and inverse Lambert's projection in Equation (\ref{EQN:LAMBERT_PROJ}). The domes in D--E were obtained using the DistMesh method \cite{REF:46} followed by rescaling and the inverse Lambert's projection. A) A mesh grid resolution of $12$ results in $36$ vertices; B) A mesh grid resolution of $20$ results in $100$ vertices; C) A mesh grid resolution of $60$ results in $900$ vertices; D) An edge element size of $0.325$ results in $35$ vertices; E) An edge element size of $0.185$ results $103$ vertices; F) An edge element size of $0.635$ results in $896$ vertices.
\label{FIG:geodesic_domes}
}
\end{figure}

We can reconstruct the surface patch from Eq.~(\ref{EQN:SERIES2}) when the coefficients are known. To obtain a homogeneous representation of all the morphological features over the cap domain, this step requires uniformly sampled points (vertices). In ordinary SHA, convex icosahedrons (geodesic polyhedrons) with different mesh refinements are often used for the reconstruction because the vertices of these icosahedrons are equally spaced. Analogously, for SCHA, we use geodesic domes with uniformly distributed vertices over a spherical cap (see Figure~\ref{FIG:geodesic_domes}). To construct these domes, we first construct a uniformly meshed unit disk and then use the inverse Lambert azimuthal equal-area projection to project it to the desired spherical cap. Like the approach presented in Section \ref{SEC:PARAMETRIZATION}, we need to rescale the unit disk by an appropriate scaling factor $r_l$ to yield a spherical cap with the correct area under the projection:
\begin{equation}
    \label{EQN:SCALE_LAMBERT}
    r_l = \sqrt{2(1-\cos\theta_c)}.
\end{equation}

\begin{figure}
\centering
\includegraphics[width=0.7\textwidth]{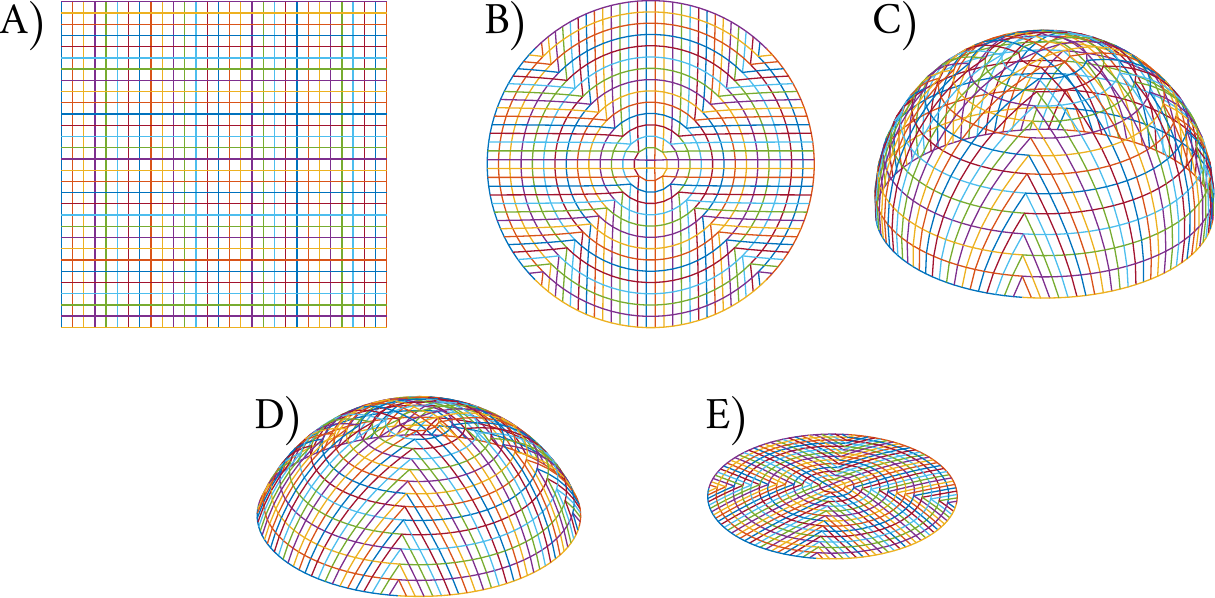}%
\caption{Transformation of a square grid to a geodesic dome grid (before triangulating the surfaces) using the algorithm proposed by Ro{\c{s}}ca (2010) \cite{REF:45}. A) A square grid with $60$ horizontal and vertical lines resulting in $900$ points of intersection (vertices); B) An area-preserving map of the square grid onto a quad-grid disk; C) The final geodesic dome with $\theta_c = \pi/2$ (a hemisphere); D) The final geodesic dome with $\theta_c = \pi/3$; E) The final geodesic dome with $\theta_c = \pi/18$.
\label{FIG:geodesic_domes_grid}
}
\end{figure}

\par
For reconstructing the surface patch, we need to choose a suitable mesh refinement for practical and theoretical aspects. Practically, the complexity of the numerical models, e.g. finite and discrete element methods, depends on the number of elements used to discretise the continuum. Theoretically, we need a mesh size that represents the smallest required detail depending on the application. If we imagine that a simple 1D wave propagates along a discrete line, we need at least five knots (points) to capture the sign changes along a full wave. This means that the distance between two adjacent vertices at a certain mesh refinement must be at least one-fourth of the target minimum wavelength. The trade-off between capturing morphological features and the numerical complexity should be calibrated through several reconstruction trials. More details about the minimum and maximum wavelength associated with the SCHA will be discussed in Section~\ref{SUBSEC:TEHTA_C}.
\par
We consider two approaches for constructing a disk with a uniform grid. The first approach, proposed by Ro{\c{s}}ca (2010) \cite{REF:45}, builds the disk-grid by transforming a rectangular grid of squares to a quad-grid disk using an area-preserving map, which can be subsequently triangulated using standard Delaunay triangulation. This approach gives us direct control of the number of vertices on the spherical cap. Another approach is the DistMesh method proposed by Persson and Strang (2004) \cite{REF:46}, in which one can construct a uniform mesh on a unit disk by a force-based smoothing approach. For this approach, the desired length of the edge elements on the unit disk is a required input that indirectly controls the number of vertices on the surface mesh. For both approaches, we need to rescale the resulting unit disk meshes using Eq.~(\ref{EQN:SCALE_LAMBERT}) (the rescaled disk is denoted by $\mathcal{D}_l$) before applying the inverse Lambert's projection $\tau^{-1}_l$ so that every $(x,y)$ point on the rescaled disk is projected to the $(X, Y, Z)$ on a spherical cap to $\mathbb{S}^2_{~\theta \leq \theta_c}$ as:
\begin{equation}
\label{EQN:LAMBERT_PROJ}
        \tau^{-1}_l(x, y) = (X,Y,Z) =\\
        \left(\sqrt{1-\frac{x^2 + y^2}{4}}x, \sqrt{1-\frac{x^2 + y^2}{4}}y,-1+\frac{x^2 + y^2}{2}\right).
\end{equation}
Figure \ref{FIG:geodesic_domes_grid} visualises the process in \cite{REF:45}, and Figure \ref{FIG:geodesic_domes} shows the final geodesic domes obtained using the two approaches.

\section{Shape descriptors and fractal dimension (FD)}
\label{SEC:FRACT}
\par
Like the traditional Elliptical Fourier Descriptors (EFD) \cite{REF:38} and their 3D extension to SH \cite{REF:1}, SCHA also yields coefficients that can describe and reconstruct surfaces. Because the SCH is a \say{local extension} of the ordinary SH, it matches most of its properties \cite{REF:44}, such as the spectral power analysis for determining the surface fractality. In this section, we describe the interpretation of the shape descriptors derived from SCHA and their relationship to those derived from other traditional methods like HSH. We also show how to estimate the wavelengths associated with each degree and extract the fractal dimension (FD) from the attenuation of the shape descriptors.

\subsection{Shape descriptors derived from SCHA}
\par
Shape descriptors tell us how similar or dissimilar surfaces are to each other and are particularly useful for matching or classifying complex topologies. These shape descriptors must be invariant to the location, rotation and size of the targeted surface. In this section, we compare the shape descriptors derived from SCHA to those derived from SHA or EFD and explain some additional properties of the SCH regional analysis.
\par
By expanding the signal $f(\theta,\phi) = (x(\theta,\phi), y(\theta,\phi), z(\theta,\phi))^{T}$, we simultaneously obtain the coefficients for the three orthogonal signals. We first consider the \say{breathing} mode (corresponding to $k=0$), where we have a constant shift for all the points in the signal. The corresponding real-valued ($k=m=0$) coefficients along the $x$, $y$ and $z$ axes are represented by ${}^{\theta_c}q^0_{0, x}$, ${}^{\theta_c}q^0_{0, y}$ and ${}^{\theta_c}q^0_{0, z}$. These coefficients correspond to the geometric centroid of the surface, which is similar to the EFD and SHA methods. Practically, we ignore these coefficients and set them to zero to make the shape descriptors and the reconstruction invariant to the location.
\par
\begin{figure}
\centering
\includegraphics[width=0.7\textwidth]{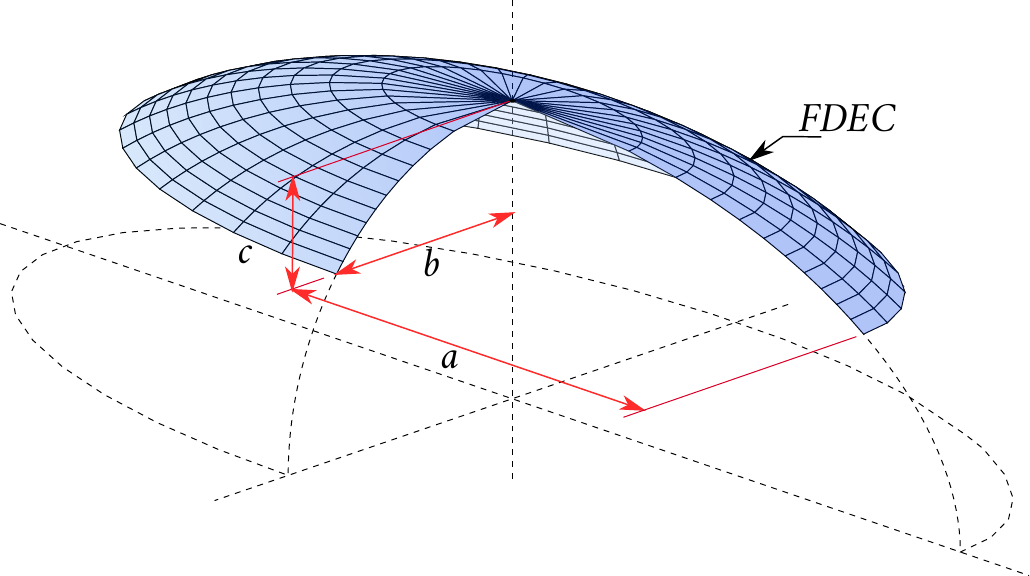}%
\caption{The first-degree ellipsoidal cap (FDEC) defined by $k = 1$; $a$ is the half-major axis, $b$ is the half-minor axis and $c$ is the ellipsoidal cap depth, where $|c| \leq |b| \leq |a|$.
\label{FIG:FDEC}
}
\end{figure}

For $k = 1$, EFD basis functions define an ellipse in $2D$, and SH functions define an ellipsoid in $3D$. However, in our case, the basis functions describe an ellipsoidal cap that is sized by the $3\times 3$ matrix ${}^{\theta_c}q^m_{1}, ~ m\in\{-1,0,1\}$, herein referred to as the first-degree ellipsoidal cap (FDEC). The FDEC determines the size and the orientation of the reconstructed surface, and it also shapes the main domain (spatial space) of the other harmonics (for all $|m|>1$). Following Brechb\"uhler et al. \cite{REF:1}, the size of the FDEC can be determined by rearranging the coefficients from $k = 1$ to correspond to the ellipsoidal equation in the Cartesian coordinates:
\begin{equation}
    \label{EQN:FDEC_MAT}
    A = \left({}^{\theta_c}q^{-1}_{1}-{}^{\theta_c}q^{1}_{1}, i({}^{\theta_c}q^{-1}_{1} + {}^{\theta_c}q^{1}_{1}), \sqrt{2}{}^{\theta_c}q^{0}_{1}\right).
\end{equation}
Solving the eigenvalue problem of $AA^{T}$ will provide three eigenvalues $|\lambda_1| \geq |\lambda_2| \geq |\lambda_3|$. Their roots $a = \sqrt{|\lambda_1|} \geq b = \sqrt{|\lambda_2|} \geq c = \sqrt{|\lambda_3|}$ correspond to the amount of \say{stretch} along the three principal axes of the ellipsoidal cap. The directions of the \say{stretches} will follow three eigenvectors; two of them are in-plane orthogonal vectors and the third is an out-of-plane perpendicular vector. These three unit vectors describe the principal directions of the ellipsoidal cap. Figure \ref{FIG:FDEC} shows the FDEC in an arbitrarily rotated Cartesian space.
\par
When the size of the ellipsoidal cap is determined, the higher frequency basis functions will be overlaid onto the domain of the FDEC. Numerically, the basis functions of the SCH are similar for different $\theta_c$'s, but they are either \say{stretched} or \say{compressed} over $\mathbb{S}^2_{~\theta \leq \theta_c}$. This constrains all the domains to the size of the FDEC, which makes the proposed method invariant to $\theta_c$. From the point of view of the basis functions, this theoretically implies that there is no difference between the SCHA and the HSHA, providing that we are using the same normalisation factor (e.g. Schmidt semi-normalised). However, the results of the analysis will depend on the introduced distortion between the real Cartesian space and the parameterisation space. This problem will be discussed in Section \ref{SEC:HALF_ANGLE_PATCH}.
\par
The FDEC size is also important for estimating the wavelengths associated with the harmonics expanded over the surfaces. In geophysics, the SCHA is used for a regional analysis over planets and moons to directly estimate the average circumference. As we do not have a closed surface that is approximately spherical, however, we use the FDEC to estimate the average circumference of a full \say{virtual} ellipse that lies at $\theta_c \equiv \pi/2$ (Greenwich line) to estimate the circumference (more details will be discussed in Section \ref{SEC:HALF_ANGLE_PATCH}).

\subsection{Shape descriptors for estimating the fractal dimension}
\par
Shape descriptors have been extensively studied in the SH literature. The most commonly used descriptors are rotation-invariant ones that do not require the registration of the surface. Based on the amplitudes calculated in the expansion stage ${}^{\theta_c}q^m_k$, we can express the shape descriptors (for instance see \cite{REF:14, REF:57, REF:56}) as follows:
\begin{equation*}
    \hat{D}_{k, x} = \sqrt{\sum_{m = -k}^{k} ||{}^{\theta_c}q^m_{k,x}||^{2}},
\end{equation*}
\begin{equation*}
    \hat{D}_{k, y} = \sqrt{\sum_{m = -k}^{k} ||{}^{\theta_c}q^m_{k,y}||^{2}},
\end{equation*}
\begin{equation*}
    \hat{D}_{k, z} = \sqrt{\sum_{m = -k}^{k} ||{}^{\theta_c}q^m_{k,z}||^{2}}.
\end{equation*}
Then, the resultant (radial) rotation-invariant descriptors can be expressed as:
\begin{equation*}
    \hat{D}_{k}^2 = \hat{D}_{k, x}^2 + \hat{D}_{k, y}^2 + \hat{D}_{k, z}^2.
\end{equation*}
These descriptors need to be normalised with the size of the FDEC to obtain size-invariant descriptors and to estimate the Hurst exponent, important for removing the dimensionality of the data \cite{REF:17}. Thus, the final shape descriptors can be written as:
\begin{equation} \label{EQN:SHAPE_DESC}
    {D}_{k}^2 =\frac{\hat{D}_{k}^2}{\hat{D}_{1}^2}, ~\forall k>1.
\end{equation}
For nominally flat and circular surface patches, where the roughness (noise) only shifts the $z-$coordinates, $\hat{D}_{k,x}^2 \approx \hat{D}_{k,y}^2 \to 0$ for $k>1$.
\par
The power spectral analysis determines the power accumulated at a certain wavelength of the analysed surface, while the spectral attenuation computes the FD of the surface. Similar to the PSD analysis, here we found that the accumulated power at each degree correlates with the Hurst exponent $H$ (see \cite{REF:17, REF:59} for details) as follows:
\begin{equation} \label{EQN:FD_PRO}
    {D}_{k}^2 \propto k^{-2H}.
\end{equation}
The FD can then be computed as (see \cite{REF:17}):
\begin{equation} \label{EQN:FD}
    FD = 3 - H.
\end{equation}
For fractal surfaces, the Hurst coefficient is between 0 and 1 ($0<H<1$), and the FD is between 2 and 3 ($2<FD<3$). The higher the $H$ exponent, the smoother the surface.
\par
The Hurst exponent can be estimated from the rate of power attenuation with an increasing $k$, as the slope of the fitted line on a log-log graph is $-1/(2H)$. One common way to estimate the exponent is through the least squares fit of the equation $D_k^2 = a k^{-2H}$, where $a$ is the proportionality constant in Eq.~(\ref{EQN:FD_PRO}). This approach estimates a radially symmetric Hurst exponent for self-affine surfaces. Unlike the Hurst exponents computed by traditional methods that investigate profiles along different directions, the Hurst exponent herein computed on the basis of SCH coefficients is invariant to directions or rotations. We benchmarked this approach against artificial fractal surfaces, and the results are presented in Section \ref{SUBSEC:RES_FD}.

\section{Choosing the optimal half-angle \texorpdfstring{$\theta_c$}{Lg} and the appropriate surface patch}
\label{SEC:HALF_ANGLE_PATCH}

\subsection{Choosing the optimal half-angle \texorpdfstring{$\theta_c$}{Lg}}
\label{SUBSEC:TEHTA_C}
\par
In Section \ref{SEC:FRACT}, we explained that the SCH basis is invariant to $\theta_c$, raising the question: What is the optimal half-angle $\theta_c$ for the analysis? To answer this, we need to consider two aspects: i) the numerical stability of the hypergeometric functions and ii)~the distortion between the object's space and the parameterisation space. The stability of the hypergeometric functions was briefly discussed in Section~\ref{SEC:SCH}. In general, the smaller the $\theta_c$, the faster the convergence (less terms needed), but $\theta_c$ should not be too close to the singularity points $\theta_c \in \{0, \pi\}$.
\par
If this restraint is considered, we can choose the half-angle $\theta_c$ such that the distortion introduced by the parameterisation is minimised. As compared to HSH, SCH offers this additional parameter $\theta_c$ to minimise the distortion. We here adopt the area distortion ($d_{area}$) used in Eq.~(\ref{EQN:AREA_DIST}) to determine the optimal $\theta_c$, though other measures can also be used.
\par
To find the optimal $\theta_c$ that minimises the area distortion, we used the PSS algorithm \cite{REF:65}, which we used already for finding the optimal M\"obius transformation for $h(X,Y)$ (see Section \ref{SEC:PARAMETRIZATION}). {Note that we used the area distortion measure $d_{area}$ as the objective function as the proposed parameterisation method is highly conformal and always yields a minimal angular distortion regardless of $\theta_c$.} We searched for the optimal $\theta_c$ in a domain limited by a lower bound $\theta_{LB}$ and an upper bound $\theta_{UB}$. The optimisation problem can then be written as:
\begin{equation}
\label{EQN:OPT_TH_C}
\begin{aligned}
\min_{\theta_c} \quad & \Big( \min_{X, Y} d_{\text{area}} \Big),\\
\textrm{s.t.} \quad & \theta_{LB} \leq \theta_c \leq \theta_{UB}.\\
\end{aligned}
\end{equation}
\par
By investigating several benchmarks, we observed that the algorithm chooses the $\theta_c$ that best represents the global concavity of the open surface. For instance, for the nominally flat surfaces examined in Section \ref{SEC:NUM_EXP}, we find that $\theta_c$ is always as low as $\theta_{LB}$, which we set to $\pi/18=10^{\circ}$. For the portion extracted from the scanned stone (see Section~\ref{SUBSEC:APP_OPT_THETA_C}), we obtain an optimal half angle of $99.9158^{\circ}$, which corresponds to an area distortion of $0.240406$. For practical reasons, we can assume $\theta_c \approx 90^{\circ}$ (a hemisphere) with an area distortion of $0.253300$. In addition to the natural free edge, this helps explain why using HSH led to better results than SH when studying skulls and ventricles in medical-imaging problems \cite{REF:18, REF:19}.

\subsection{Wavelength analysis with \texorpdfstring{$\theta_c$}{Lg}}
\par
The literature review in Section \ref{SUBSEC:LIT_SPH} highlighted that separating the shape of an object from its roughness is a challenging problem, with most prior works proposing the use of empirical rules based on the reconstruction quality. We here offer an alternative approach adopted from the geophysics literature on SCH that determines the required number of degrees based on the targeted wavelength range.
\par
Bullard (1967) \cite{REF:43} defined the wavelength in SH by $\omega = 2 \pi r/n$, where $n$ is the SH degree and $r$ is the average radius of the sphere. Haines (1988)~\cite{REF:24} derived the relationship between the degree and the spherical cap half-angle $\theta_c$ using an asymptotic approximation  (for large $k$ and small $m$):
\begin{equation}
    l(m)_k \approx \frac{\pi}{2 \theta_c} \Bigg(k + \frac{1}{2}\Bigg) - \frac{1}{2},
    \label{EQN:WAV_HAISNE1}
\end{equation}
where $\theta_c$ is in radians. Analogous to \cite{REF:43}, Haines concluded that the minimum spherical cap wavelength associated with the ${k}_{\max}$ is $\omega_{\min} = 2 \pi r/{l_{(m)_{k,\max}}}$. Solving Eq.~(\ref{EQN:WAV_HAISNE1}) for ${k}_{\max}$ gives:
\begin{equation}
    {k}_{\max} \approx \frac{2 \theta_c}{\pi} \Bigg(\frac{2 \pi r}{\omega_{\min}} + \frac{1}{2}\Bigg) - \frac{1}{2},
    \label{EQN:WAV_HAISNE2}
\end{equation}
where $k_{\max}$ is the maximum index required to cover or represent a minimum wavelength on the expanded cap.
\par
In geophysics, the term $2\pi r$ is the average circumference of the earth (or any other planet under consideration). As we mentioned earlier and unlike in geophysics, here we expand three orthogonal signals instead of the radial one. The size of the domain is described by the three parameters $a$, $b$ and $c$, which are determined by the FDEC for $k = 1$.
\par
Bullard (1967) \cite{REF:43} derived this expression from the number of divisions along the circumference for a certain degree $n$; the maximum circle defined over the sphere (equator) occurs when $m = n$ (sectoral harmonics). Furthermore, he proved that this is true at any point on the sphere for any arbitrary pole. Here, we apply the same concept by replacing $n$ with $k$ (the index of the degrees). At $m = k$, the largest ellipse defined over the spherical cap changes its sign $k$ times. Thus, $\omega = \zeta/k$, where $\zeta$ is the circumference of the largest ellipse defined over the first degree ellipsoidal cap (FDEC) (Figure \ref{FIG:FDEC}). Additionally, $\zeta$ can be estimated from the FDEC, which is described by the orthogonal vectors ${}^{\theta_c}\vec{q~}^{-1}_{1}$, ${}^{\theta_c}\vec{q~}^0_{1}$ and ${}^{\theta_c}\vec{q~}^1_{1}$ (Section \ref{SEC:FRACT}). The wavelength at an index $k$ can then be written as:
\begin{equation}
    \label{EQN:WAVELENGTH}
    \omega_k \approx \frac{2\pi}{k}\sqrt{\frac{a^2 + b^2}{2}}, \forall k \geq 1.
\end{equation}
\par
Equation (\ref{EQN:WAVELENGTH}) can also be reached by assuming that the largest ellipse on an ellipsoidal cap corresponds to the equator (runs through the Greenwich line), and thus $\theta_c \to \pi/2$. Then, by letting $r \approx \sqrt{(a^2 + b^2)/2}$, the asymptotic formula in Eq.~(\ref{EQN:WAV_HAISNE2}) will be the same as in Eq.~(\ref{EQN:WAVELENGTH}). Note that the ($\approx$) sign in the above expressions denote that these formulae have been derived asymptotically~\cite{REF:21}, the circumference of the largest ellipse is also approximated, and the circumference of the FDEC is not the final circumference of the expanded surface patch. This equation is more accurate for circular patches where the FDEC perimeter is more circular than elliptical.

\subsection{Choosing the appropriate surface patch}
\par
Choosing the appropriate sampled surface patch eases the parameterisation process. It also avoids over-/under-representation of certain morphological features over the targeted surface, accordingly preventing potential problems in the subsequent analysis step. The following points are recommendations for sampling a surface patch:

\begin{itemize}
    \item Manifold edges and surfaces without opening (genus-0) meshes are required for the parameterisation algorithm proposed in this paper.
    \item The input mesh must not contain duplicated faces, edges, vertices or skewed faces with close-to-zero areas.
    \item A round surface patch avoids point clusters at sharp corners as a result of the parameterisation algorithm, which would lead to a poorly represented morphology. 
    \item Circular patches with no sharp edges will cause the first ellipse to be mostly circular. As a result, the wavelength definition will be more accurate because the circumference estimate is more precise.
    \item The scanned surface patch should be sampled as uniformly as possible, which will make the parameterisation algorithm more accurate due to the use of the least-squares fitting method for computing the coefficients of the SCHA.
    \item When the surface patch does not have enough vertices to compute all degrees (results in badly conditioned matrix $B$ in Eq.~[\ref{EQN:LSF}]), the surface should be subdivided as needed to increase the number of expanded equations and increase the quality of the coefficients.
    \item When the results are wavy when reconstructed with high degrees, we recommend subdividing the overall surface or the local wavy areas to increase the number of expanded equations, thus reducing the fitting error.
\end{itemize}

\section{Windowing the spectral domain and roughness projection}
\label{SEC:ROUGHPROJ}
\par
In this section, we introduce a method for projecting the roughness on surfaces using the analysis results expanded from the SCH, HSH and ordinary SH methods. This is useful for studies of contact problems, as it allows for systematically modifying the bandwidth of wavelengths included in the reconstruction and for altering a surface geometry. The isolated or generated surface roughness can then be projected onto a donor mesh of the user's choice. 
\par
Let us assume that we have obtained the SCH coefficients up to the order $K_{\max}$ either through expansion or through artificial generation (not included in this paper). The rough surface is to be projected onto a mesh that donates the general shape of the object, henceforth called the \say{donor mesh}. The targeted vertices on the donor mesh can be determined and parameterised to obtain $f_{d}(\theta,\phi) = \big(x_{d}(\theta,\phi), y_{d}(\theta,\phi), z_{d}(\theta,\phi)\big)^{T}$.
\par
For the SH, we combined the parameterisation approaches by Choi et al. in 2015 \cite{REF:49} and 2020 \cite{REF:50} to obtain a conformal spherical parameterisation of the donor mesh with minimal area distortion. For the HSH and the SCH, we used the approach proposed in Section \ref{SEC:PARAMETRIZATION}. However, instead of parameterising only the spherical cap, we parameterised the whole donor mesh over a unit sphere and then constrained the targeted domain of the vertices to a spherical cap with $\theta_c$.
\par
After parameterising the targeted vertices over the donor mesh, we projected the roughness onto the surface using a simple bandwidth-limiting (windowing) function in the spectral domain. Let us assume that $\bar{k}_{\min} \geq 1$ and $\bar{k}_{\max} \leq K_{\max}$ are the minimum and maximum indices, which correspond to $\omega_{{\max}}$ and $\omega_{{\min}}$ wavelengths computed from Eq.~(\ref{EQN:WAVELENGTH}). The roughness projection can then be implemented by modifying Eq.~(\ref{EQN:SERIES2}) to:
\begin{equation}
    f_{d}(\theta, \phi) \approx \sum_{k = \bar{k}_{\min}}^{\bar{k}_{\max}} \sum_{m = -k}^{k} {}^{\theta_c} q^m_k~ {}^{\theta_c} C_{k,d}^{m} (\theta, \phi),
    \label{EQN:ROUGH_PRO}
\end{equation}
where $C_{k,d}^{m} (\theta, \phi)$ are the basis functions evaluated for the targeted domain on the donor mesh. This expression is equivalent to multiplying the reconstruction expression in Eq.~(\ref{EQN:SERIES2}) by the traditional rectangular eigenfunction window $H(k)$ in the spectral domain (a sharp cut-off in the spectral domain) where:
\begin{equation*}
    f_{d}(\theta, \phi) \approx \sum_{k = 0}^{K_{\max}} \sum_{m = -k}^{k} H(k)~ {}^{\theta_c} q^m_k~ {}^{\theta_c} C_{k,d}^{m} (\theta, \phi),
    \label{EQN:ROUGH_PRO2}
\end{equation*}
and
\[  H(k) =  \left\{
\begin{array}{ll}
      1, & \bar{k}_{\min}\leq k \leq \bar{k}_{\max}, \\
      0, & \text{otherwise.} \\
\end{array} \label{EQN:ROUGH_PRO3}
\right. \]
\par
The coefficients can be manipulated in the spectral domain by rotating and scaling as needed. In this paper, we will not demonstrate the coefficients manipulation methods or the random generation of fractal surfaces with a certain Hurst exponent.

\section{Numerical examples and discussions for SCHA}
\label{SEC:NUM_EXP}
\par
In this section, we present and discuss the results of the proposed SCHA method. We begin with a visual benchmark that helps build the intuitive arguments we raise about the scalability and wavelengths. Second, we benchmark the SCHA against various random fractal surfaces to estimate the Hurst exponent. Then, we analyse and study a rough surface patch of a rock whose geometry was obtained by laser scanning. Finally, we demonstrate an example of a roughness projection from the scanned stone to alter the microstructure of a donor mesh.

\subsection{Visual benchmark}
\par
\begin{figure*}
\centering
\includegraphics[width=1.0\textwidth]{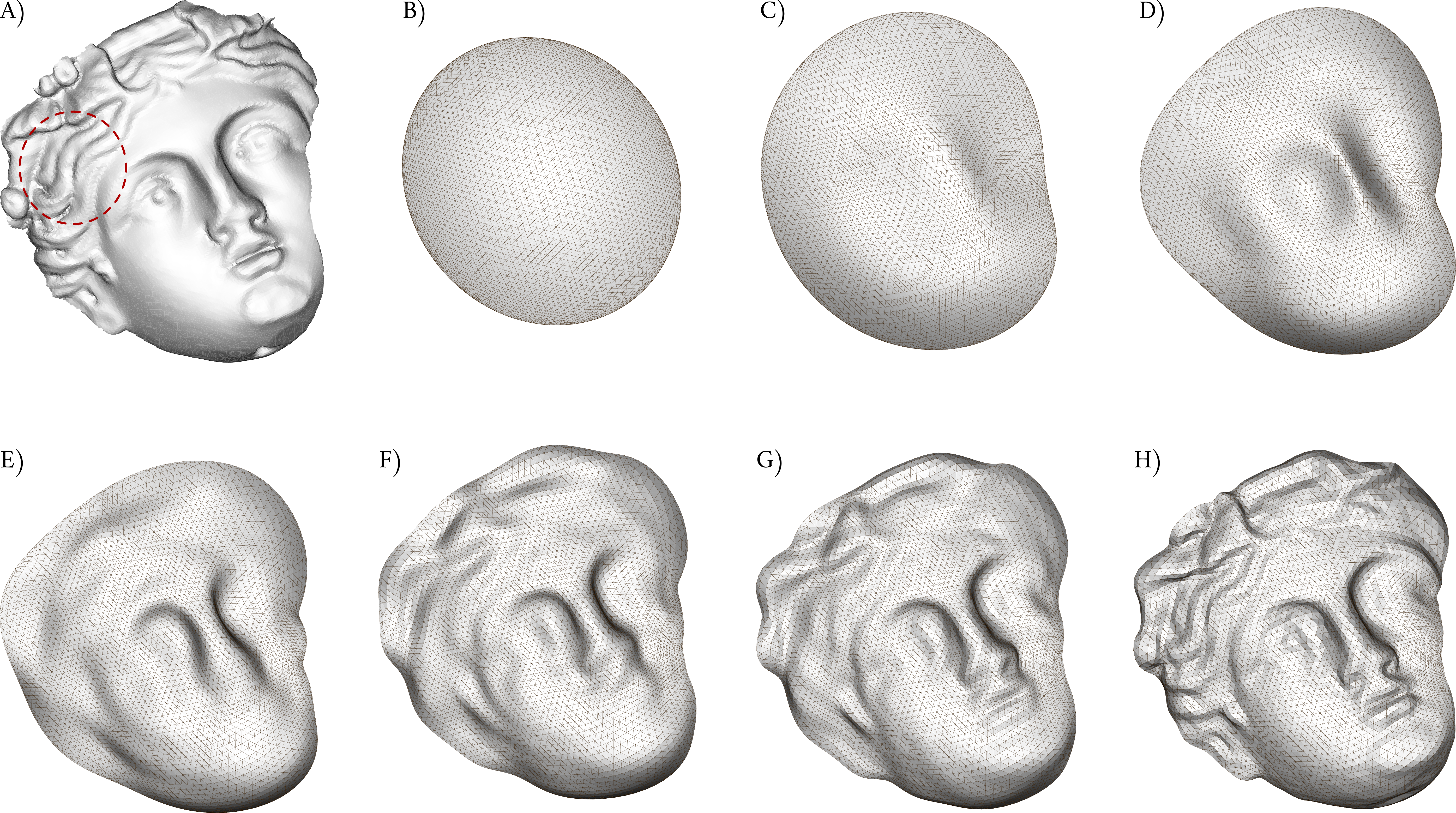}%
\caption{The reconstruction of the face sculpture with SCH. A) The input surface after remeshing with $21,280$ vertices. It was expanded up to $k = 40$ and reconstructed with an edge length of $0.032$ on a unit disk; B) The reconstruction with $k=1$ results in an FDEC with $2a = 123.381, 2b= 111.336 \text{ and } c = 19.844$; C) Reconstruction at $k = 3$; D) Reconstruction at $k = 5$; E) Reconstruction at $k = 10$; F) Reconstruction at $k = 15$; G) Reconstruction at $k = 20$; H) Reconstruction at $k = 40$.
\label{FIG:FACE_RES_Ks}
}
\end{figure*}

We use a sculpture of a face (retrieved from \cite{REF:48}) for visualising the reconstruction evolution with $k$. The 3D-scanned geometry of the face was first cleaned and remeshed to a manifold surface with a reasonable number of vertices that sufficiently describe the details. Figure \ref{FIG:FACE_RES_Ks}(A) shows the input geometry that was used. This benchmark was chosen for the following reasons:
\begin{itemize}
    \item It is a nominally flat surface that contains large and small wavelengths (facial features). To capture all the small details, a high number of orders is required.
    \item It has a significant number of vertices (more than $20,000$), which increases the problem complexity.
\end{itemize}
\par
In this example, we used $\theta_c = \pi/18$ to parameterise, expand and reconstruct the surface, as this half angle is low enough to minimise the area distortion but also high enough for the hypergeometric functions to be stable. The results in Figure \ref{FIG:FACE_RES_Ks}(B)--(H) show the evolution of the reconstruction stages with $k$.
\par
\begin{figure}
\centering
\includegraphics[width=0.42\textwidth]{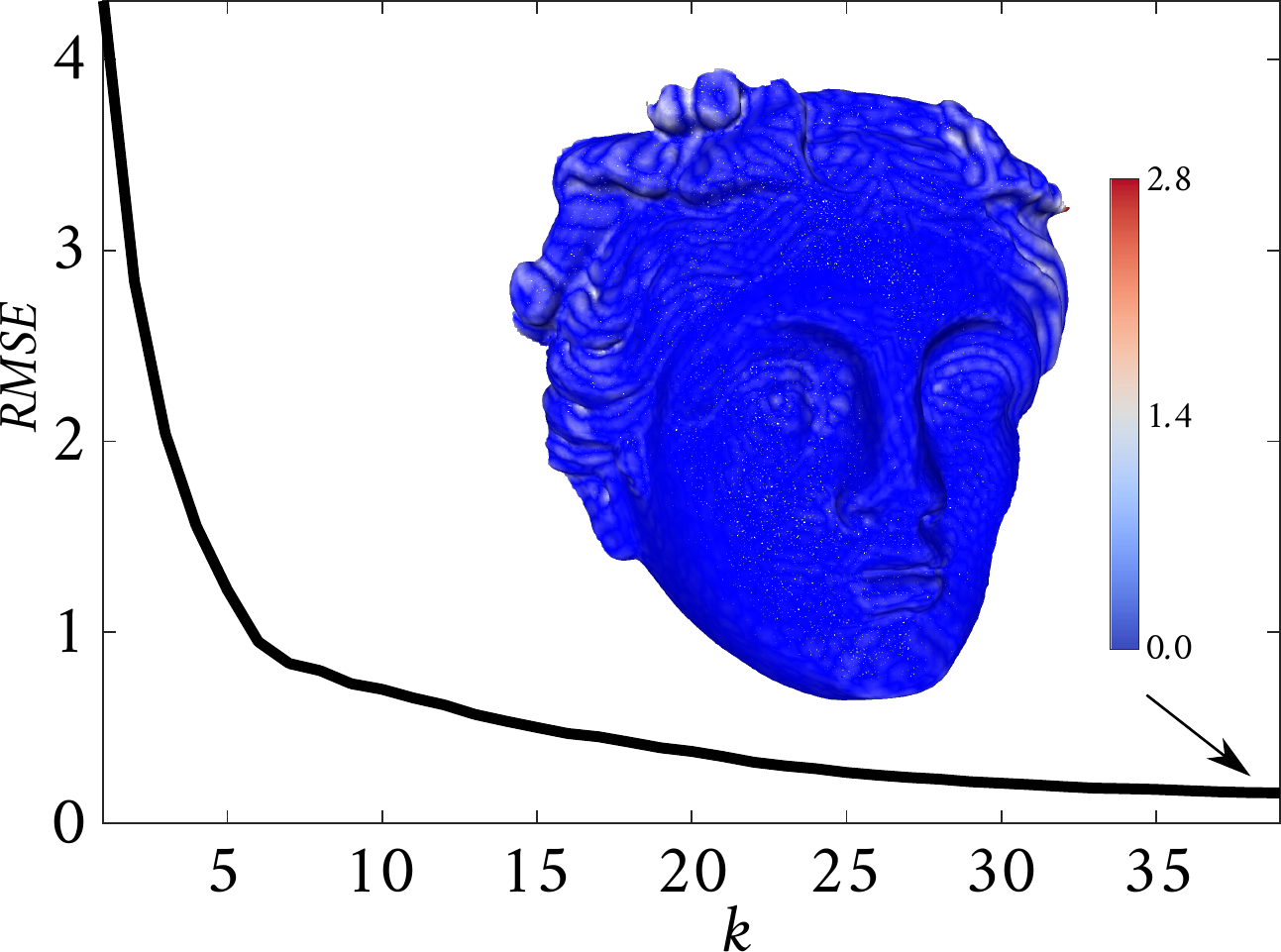}%
\caption{The root mean square error (RMSE) for the face sculpture. The inset shows the Hausdorff distance at $k = 40$, where $1,221,276$ points were sampled on the original mesh and were compared with the reconstructed mesh. For $\theta_c = \pi/18$, the RMSE for the Hausdorff distance was $0.229886$, the mean error was $0.141412$ and the highest error value scored was $2.799484$. The reported data was extracted with the open source package MeshLab \protect\cite{REF:47}.
\label{FIG:FACE_RES_RMSE}
}
\end{figure}

The method starts with fitting the large wavelengths (low $k$'s). The large facial features, such as the nose bone, begin to take shape at $k=3$. The eyebrow ridges begin to take shape at $k = 10$. Fine details like the mouth and the hair curls become visible only with higher $k$'s. One way to measure the convergence of the spatial details is by simply measuring the distance between the sampled points (Monte Carlo) on the input surface and the closest equivalent ones on the reconstructed surface at a certain $k$. The root mean square error (RMSE) of these points is then calculated using MeshLab \cite{REF:47} as the Euclidean distance normalised by the diagonal of the bounding box of the surface. The results are shown in Figure \ref{FIG:FACE_RES_RMSE}; for $k=40$, the RMSE averaged over all points is $0.158282$. Another measurement that can compare the reconstruction of the original shape is the Hausdorff distance; see the inset included in the same figure. It should be mentioned that when $k = 40$, the largest captured feature varied at a wavelength of $184.589$ mm, and the smallest varied at $4.858$ mm.
\par
Additionally, to investigate the influence of $\theta_c$ on the quality of the reconstruction, we also expanded the input mesh up to $k = 40$ assuming $\theta_c = 5\pi/18$. The RMSE for an expansion and reconstruction with $\theta_c = 5\pi/18$ was only $0.1215\%$ larger than the one with $\theta_c = \pi/18$. Practically, this means that the analysis is invariant to $\theta_c$, with the small difference due to the numerical accuracy of the hypergeometric functions and the area distortion introduced by the parameterisation. The area distortion after finding the optimal M\"obius transformation was $0.377619$ for $\theta_c = 5\pi/18$ and $0.341917$ for $\theta_c = \pi/18$.


\subsection{Wavelength and patch size}
\par
As the face sculpture example contains large and small wavelengths, high orders of SCH were needed to reconstruct the small details; compare Figure~\ref{FIG:FACE_RES_Ks}(G) and (H), where $20$ additional orders were added to capture the hair curls, the mouth and the nasal openings. To demonstrate that the number of orders needed for reconstructing a certain detail depends on the size of the patch, we considered a regional patch from the hair details to filter out the larger wavelength (facial features, e.g. the nose). The considered local patch is the red-dotted region circled on Figure \ref{FIG:FACE_RES_RMSE}(A).
\par
\begin{figure}
\centering
\includegraphics[width=0.43\textwidth]{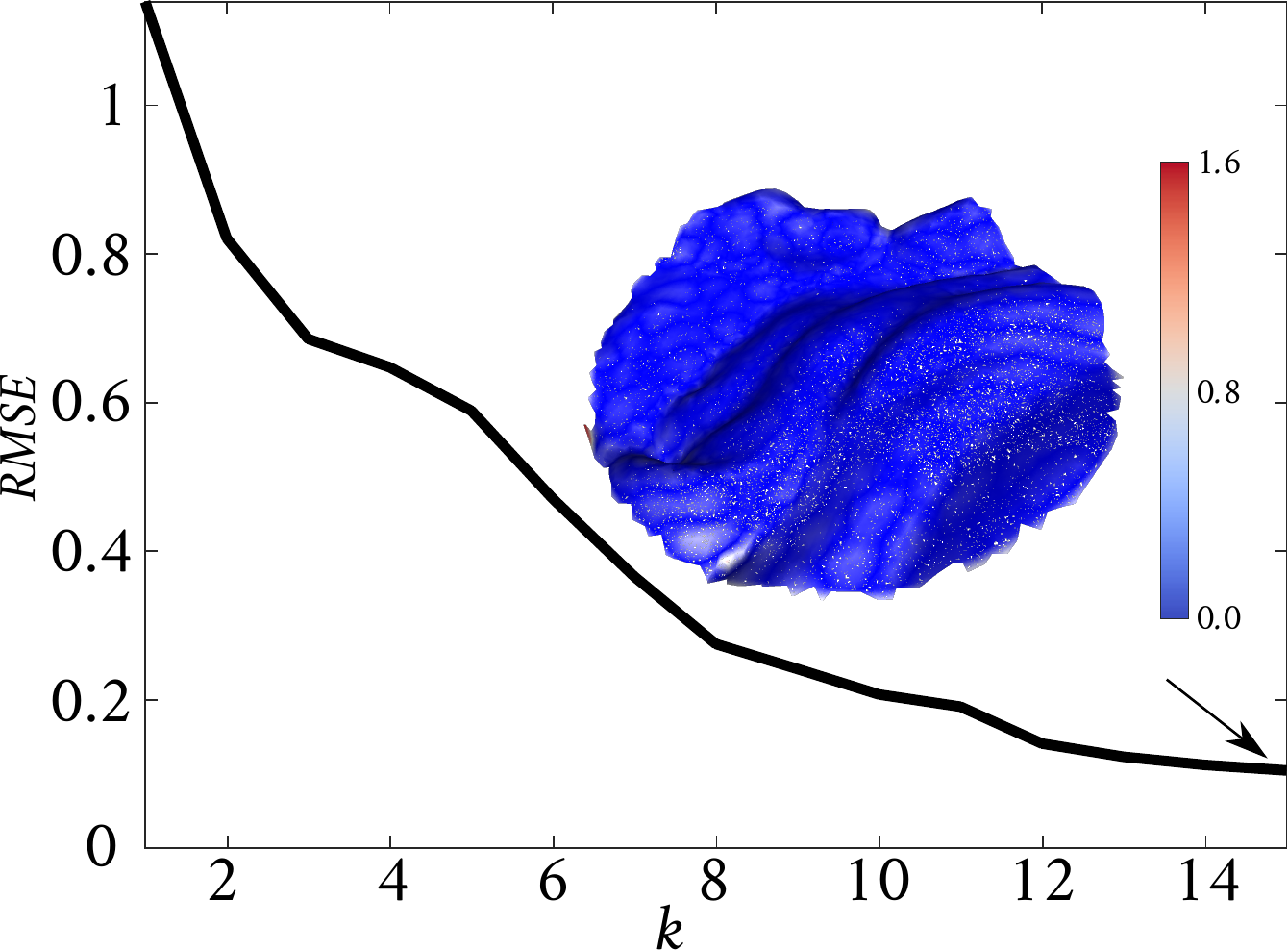}%
\caption{The root mean square error (RMSE) for the local patch on the visual benchmark. The inset shows the Hausdorff distance at $k = 15$ where we used $602,087$ points sampled on the original mesh and compared these with the reconstructed mesh. At $\theta_c = \pi/18$, the RMSE for the Hausdorff distance was $0.149861$, the mean error was $0.087238$ and the highest error value scored was $1.594269$. The reported data was extracted with the open source package MeshLab \protect\cite{REF:47}.
\label{FIG:FACE_PATCH_RES_RMSE}
}
\end{figure}

We expanded the local patch by assuming $\theta_c = \pi/18$ and using only $k = 15$. As we see from Figure \ref{FIG:FACE_PATCH_RES_RMSE}, $15$ degrees were enough to reach almost the same RMSE accuracy as the whole face with $k=40$. This resembles about $15.23\%$ of the overall computed SCH basis functions, though this ratio does not scale linearly with the computational time. The minimum RMSE for the simple distance between the surfaces at $k = 15$ was $0.104812$. By changing the range of targeted wavelengths, we can clearly see the faster convergence of this method with relatively close scales of detail (wavelengths). At only $k = 15$, the largest captured feature varied at a wavelength of $59.761$ mm and the smallest varied at $4.269$ mm, in comparison with $k = 40$, with $4.858$ mm for the full sculpture. Furthermore, we compared this with the HSHA, and we show that the error at least doubles for the different measures; see Section \ref{SUBSEC:APP_HSH} for full details.

\subsection{Analysis of rough fractal surfaces}
\label{SUBSEC:RES_FD}
\par
Here, we analysed rough artificially generated fractal surfaces, computed the Hurst coefficient from the SCHA for these surfaces and compared the obtained coefficients to the input values used for the generation. Many libraries have been developed for generating such surfaces using the traditional power spectral density (PSD) method (reviewed in \cite{REF:53, REF:55}), such as Tamaas Library \cite{REF:52}. These benchmarks are important in contact mechanics, which usually depends on exact asperity shapes, distributions and their effects on the real contact area, as per Hyun et al. (2004) \cite{REF:54}.
\par
\begin{figure*}
\centering
\includegraphics[width=1.0\textwidth]{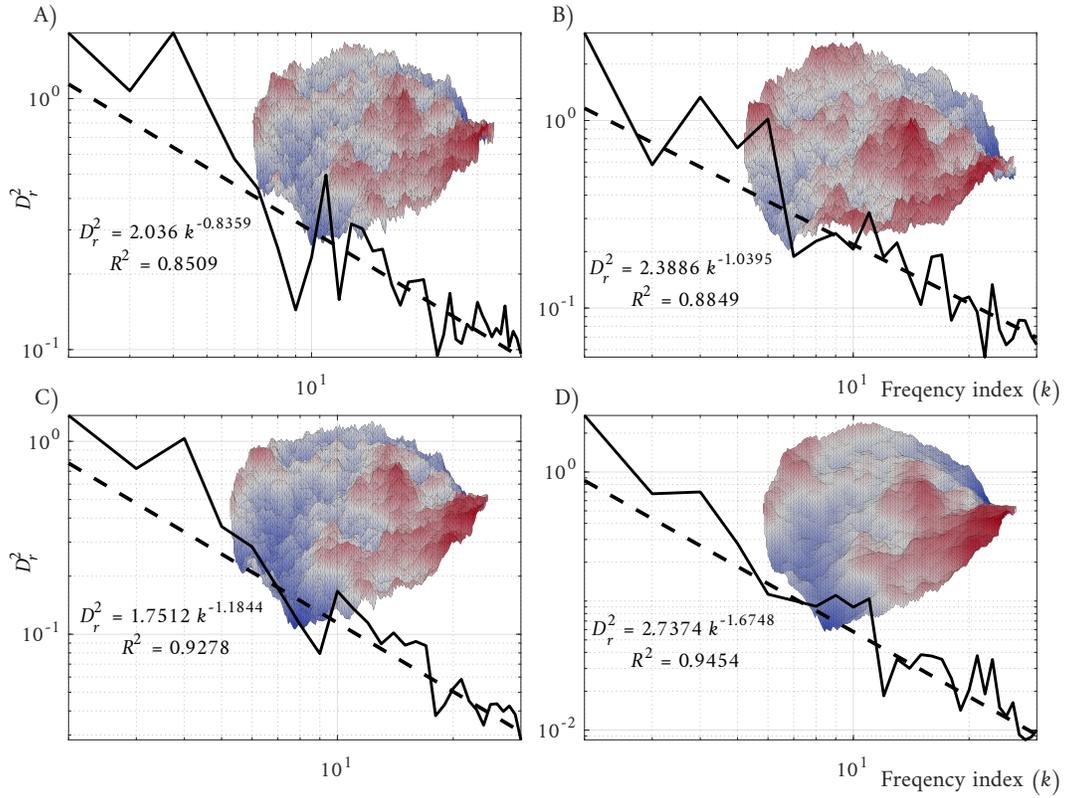}%
\caption{The analysis results for artificially generated fractal surfaces with different Hurst exponents. The solid lines represent the descriptors in Eq.~(\ref{EQN:FD_PRO}) and the dashed lines represent the fit of Eq.~(\ref{EQN:FD_PRO}). (A) a surface generated with $H = 0.4$; (B) a surface generated with $H = 0.5$; (C) surface generated with $H = 0.6$; (D) surface generated with $H = 0.9$.
\label{FIG:FRACT_RES}
}
\end{figure*}

We generated four fractal surfaces with a root-mean-squared roughness of $0.7E{-2}$ mm and Hurst exponents of $H\in \{0.4, 0.5, 0.6, 0.9\}$; see Figure \ref{FIG:FRACT_RES}. To avoid biasing the first few $k$'s on fitting the shape of the boundary lines and to expand the surfaces with fewer $k$s, notice that the generated samples are all disks (circular boundary lines). To ensure reproducibility, all the surfaces were generated with a fixed seed number of $10$ and with no roll-off wave vector. In the same figure, we show the analysis results (up to $k = 40$) in terms of the shape descriptors (solid lines) in Eq.~(\ref{EQN:SHAPE_DESC}) and the fit of Eq.~(\ref{EQN:FD_PRO}) (dashed lines).
\par
The surface in Figure \ref{FIG:FRACT_RES}(A) was generated with a Hurst exponent of $0.4$, and we estimated a Hurst exponent of $H = 0.4175$, which corresponds to a fractal dimension of $2.5825$ and an error of $+4.4\%$. For the generated surface with $H = 0.5$ (Figure~\ref{FIG:FRACT_RES}(B)), we estimated $H = 0.5197$ ($+3.9\%$) and $FD = 2.4802$. For the third example with $H = 0.6$ (Figure \ref{FIG:FRACT_RES}(C)), we estimated the $H = 0.5922$ ($-1.3\%$) and $FD = 2.4078$. However, for the fourth sample generated with $H = 0.9$ (Figure \ref{FIG:FRACT_RES}(D)), we obtained $H = 0.8370$ ($-7.0\%$) and $FD = 2.1630$. This error is typical for higher values of $H$, as the surface seems finer and most of the spatial details are part of smaller wavelengths, meaning that additional $k$s are needed to capture these finer details. Making use of the scalability of the SCHA together with the nature of the fractal surfaces (repeats itself at all scales), we cut out a smaller area at the centre of the patch in Figure \ref{FIG:FRACT_RES}(D). We analysed the latter with $k = 8$, and the estimated Hurst exponent was $0.8850$ and $FD = 2.1150$. This agrees with the conclusions drawn by Florindo and Bruno (2011)~\cite{REF:60}, who found using the PSD that the attenuation slope of $FD$ changes with the value of $k$ and that most of the fractal information is contained in the microscopic wavelengths when the change in the slope is negligible.

\subsection{Laser-scanned rough surface patch}
\par
\begin{figure}
\centering
\includegraphics[width=0.6\textwidth]{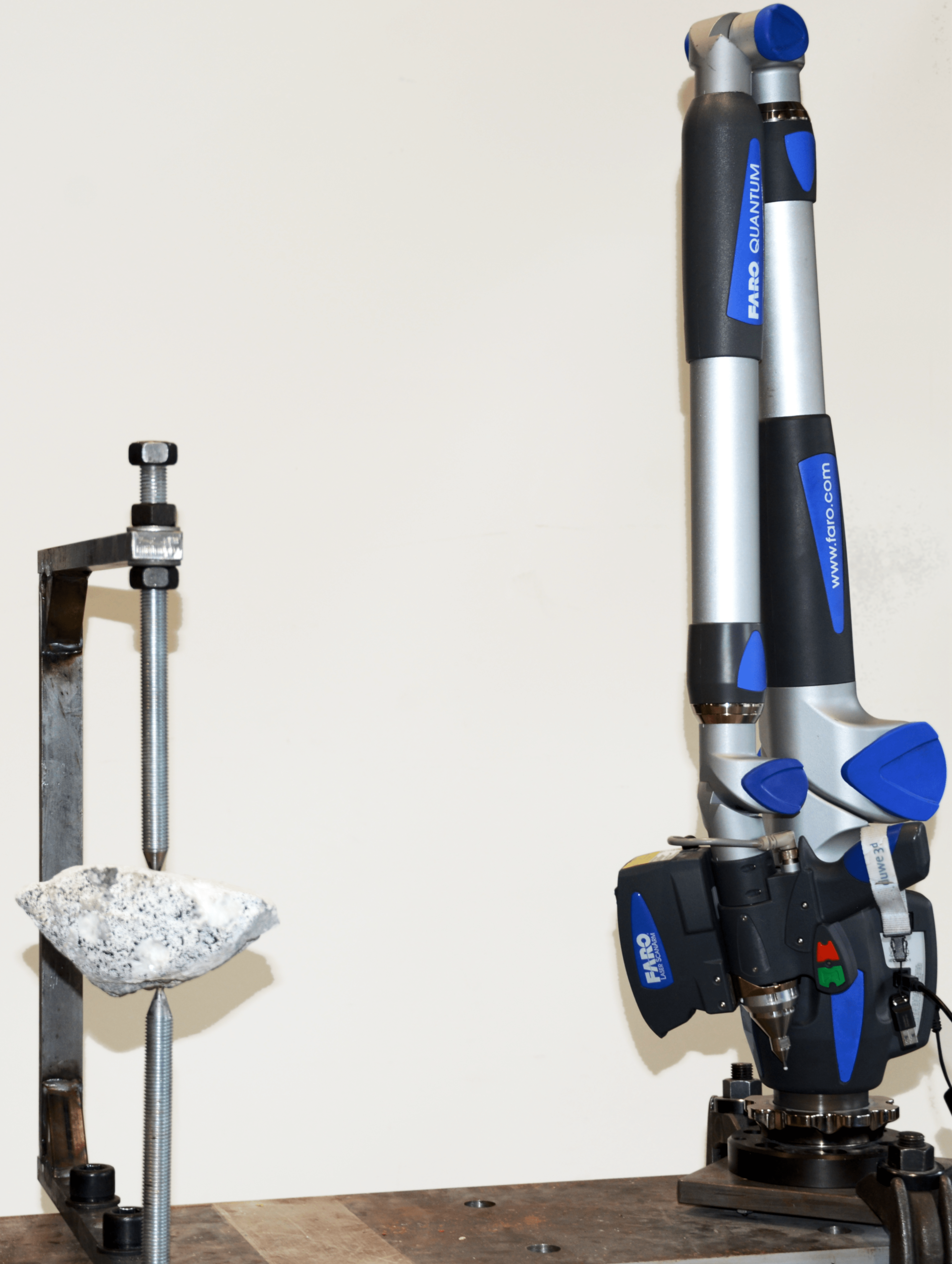}%
\caption{The laser scanner setup using the \href{https://www.faro.com/de-DE/Products/Hardware/Quantum-FaroArms}{Quantum FARO Arm$^\text{\textregistered}$}; the left side of the picture shows the laser scanner used to scan the stone, and the right side shows the stone pinned on a steel frame for the scan to minimise the loss of surface data.
\label{FIG:LASERSCANNER}
}
\end{figure}

Here, we scanned a real stone that had been previously used in an experimental campaign \cite{REF:51}. This type of stone is common in historical stone masonry walls, so studying the stones' interfacial properties helps define the mechanical behaviour of the walls. The stone surface was acquired by laser scanning (Figure~\ref{FIG:LASERSCANNER}) with an accuracy of $0.01$ mm (Figure~\ref{FIG:LS_STONE_RESULTS}). Additionally, we sampled (extracted) a small patch over the scanned surface of the stone, see Figure \ref{FIG:LS_STONE_RESULTS}, to study the roughness of the surface.
\par
\begin{figure}
\centering
\includegraphics[width=0.6\textwidth]{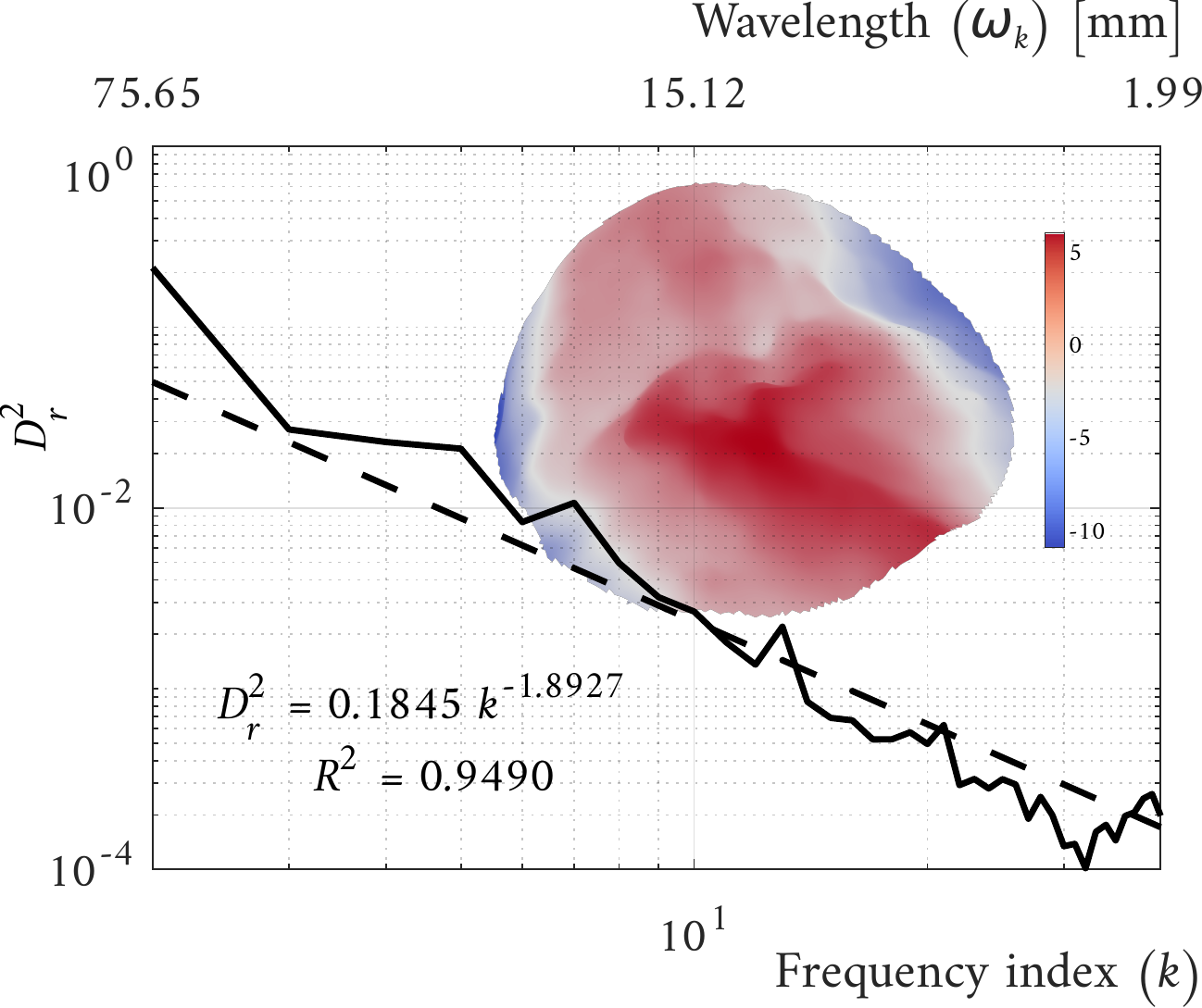}%
\caption{The SCHA for a patch sampled over the stone in Figure \ref{FIG:LS_STONE_RESULTS}. The analysis of the results for $k = 40$ and the best fit for the first $k = 12$ are shown, with $H = 0.9464$ and $FD = 2.0537$. The colour bar shows the height map of the patch in mm. The upper x-axis shows the corresponding wavelengths $\omega_{k}$.
\label{FIG:RES_REAL_PATCH}
}
\end{figure}

The surface patch was expanded up to $k = 40$. The estimated dimensions of the FDEC (computed from Eq.~(\ref{EQN:FDEC_MAT})) were: $a = 24.7879$ mm, $b = 23.3502$ mm and $c = 5.7899$ mm. From FDEC, we can estimate the wavelengths of any index $k$ from Eq.~(\ref{EQN:WAVELENGTH}). For instance, with an estimated average FDEC circumference of $151.2980$~mm, the maximum and minimum computed wavelengths $\omega_2 \approx 75.6490$ mm and $\omega_{40} \approx 1.9908$~mm, respectively. Figure \ref{FIG:RES_REAL_PATCH} summarises the results of the analysis of the descriptors. {From the analysis, the decay of the shape descriptors gives us a Hurst exponent of $H = 0.9464$, which suggests that the surface is not very rough. Later, this roughness information will be used for modifying the microstructure of a selected object.}

\subsection{Example of roughness projection}
\par
\begin{figure}
\centering
\includegraphics[width=0.40\textwidth]{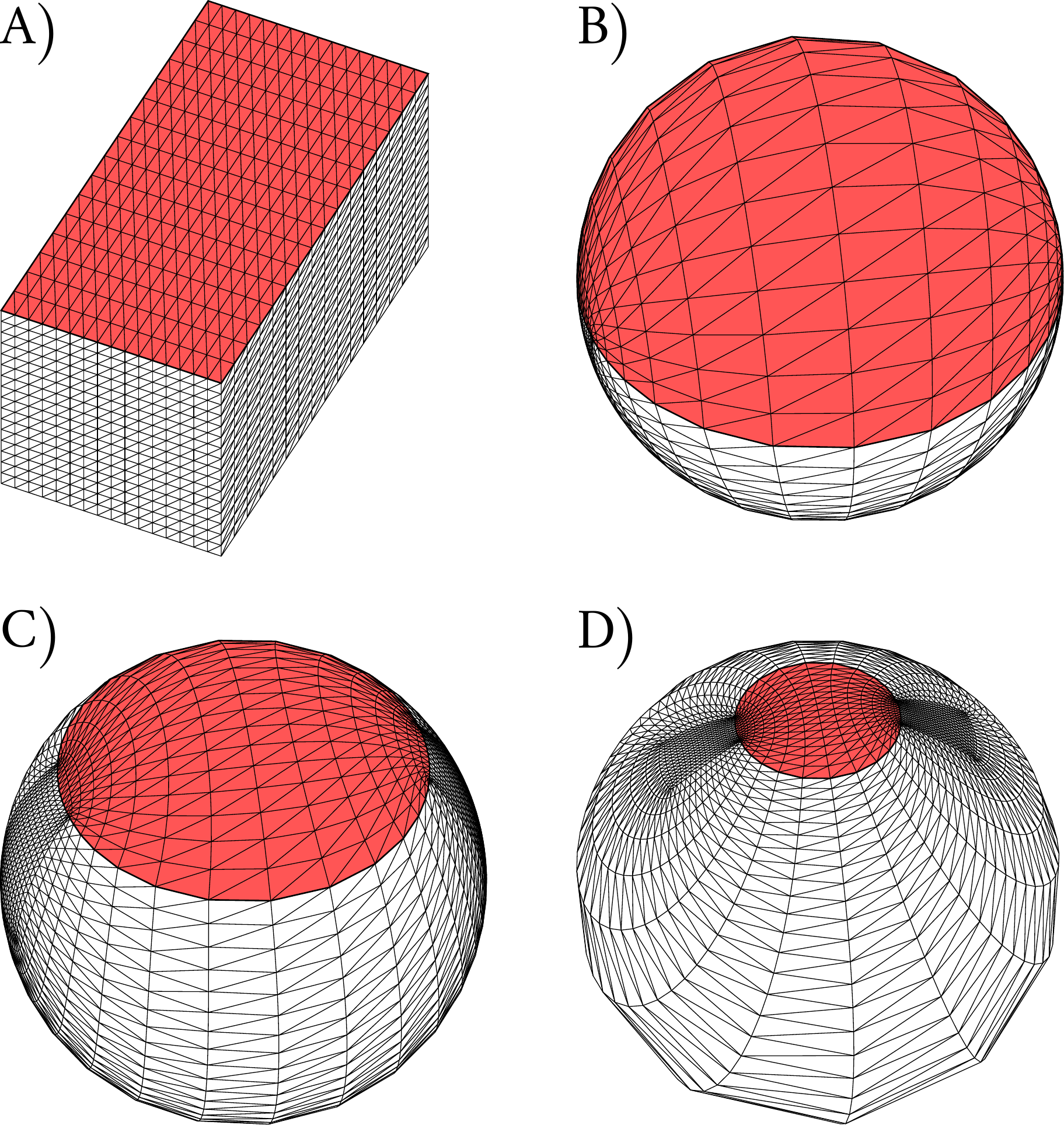}%
\caption{The parameterisation of a selected patch on a donor mesh over a unit sphere with a prescribed $\theta_c$. (A) The donor mesh with the selected (red-shaded) patch for projecting the roughness; B) $\mathbb{S}^2 \cup \mathbb{S}^2_{~\theta \leq \pi/2}$; C) $\mathbb{S}^2 \cup \mathbb{S}^2_{~\theta \leq 5\pi/18}$; D) $\mathbb{S}^2 \cup \mathbb{S}^2_{~\theta \leq \pi/9}$.
}
\label{FIG:PROJ_PARA}
\end{figure}
\begin{figure}
\centering
\includegraphics[width=0.39\textwidth]{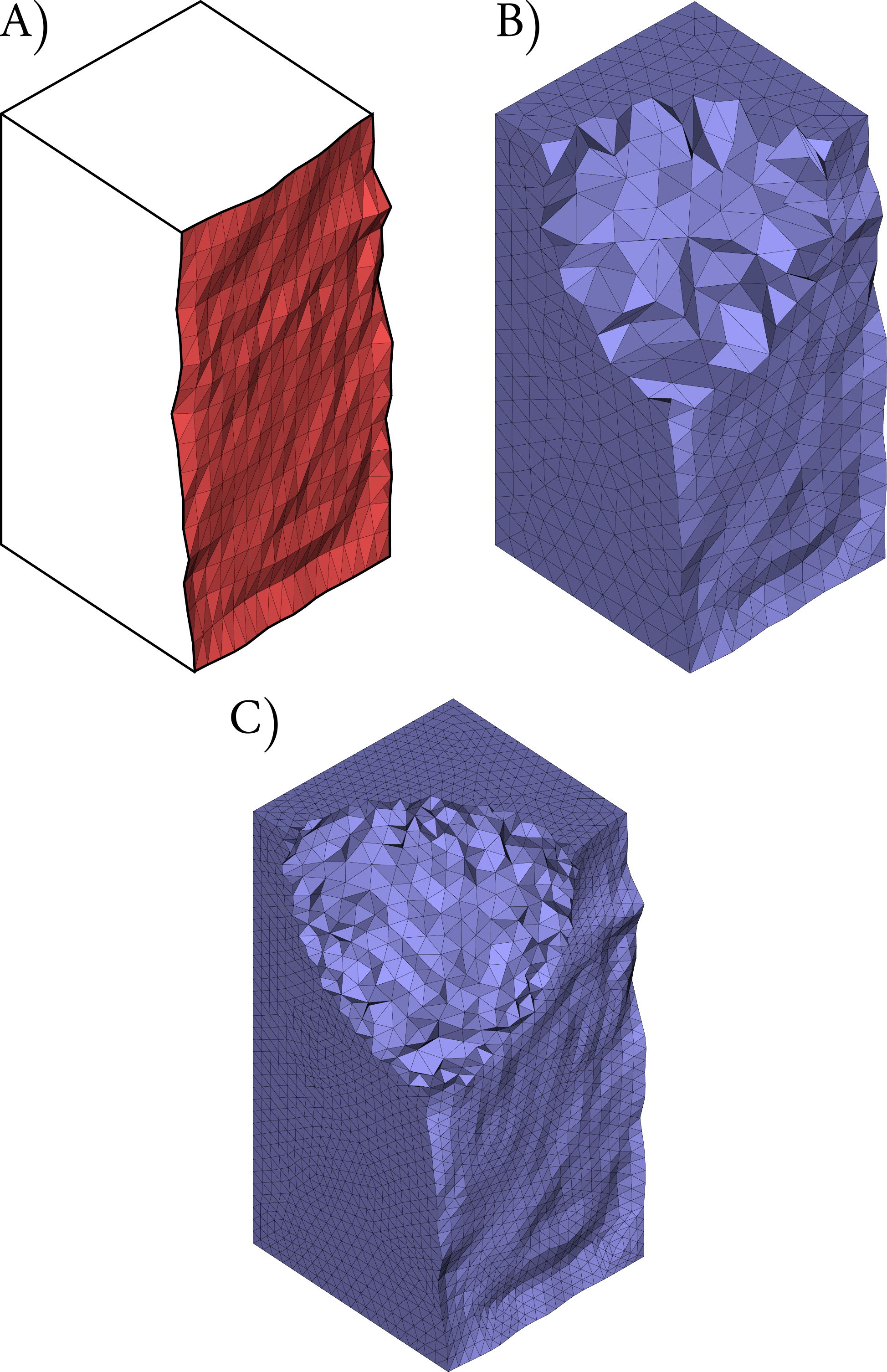}%
\caption{Roughness projected on the selected face shown in Figure \ref{FIG:PROJ_PARA}. (A) The roughness projection on the triangulated face (STL file); (B), (C) The remeshed STL surface with different refinements performed using the OpenCASCADE \cite{REF:61} library followed by volumetric finite element meshes generated by GMESH \cite{REF:62}.
}
\label{FIG:PROJ_PARA_MSH}
\end{figure}

As explained in Section \ref{SEC:HALF_ANGLE_PATCH}, we start the roughness projection by parameterising the patch on the donor mesh over a sphere with a prescribed $\theta_c$. Figure \ref{FIG:PROJ_PARA} shows the parameterisation of a selected patch on a donor mesh with various half-angles $\theta_c$. For the projection, we used the coefficients extracted from the patch in Figure \ref{FIG:RES_REAL_PATCH}. To alter the microstructure of the selected face on the donor mesh, we only include wavelengths between $w_{\max} = 15.12$ mm ($k_{\min} = 10$) and $w_{\min} = 1.99$ mm ($k_{\max} = 40$). The resulting microstructure (see Eq.~(\ref{EQN:ROUGH_PRO})) is shown in Figure \ref{FIG:PROJ_PARA_MSH}(A). To obtain the final finite element mesh, we remeshed the STL surface using the OpenCASCADE library \cite{REF:61} with various refinements (see Figure \ref{FIG:PROJ_PARA_MSH}(B) and (C)). The final volumetric finite element mesh was obtained using GMESH \cite{REF:62}.

\section{Conclusions}
\label{SEC:CON}
\par
In this paper, we conducted morphological analyses of open and nominally flat rough surfaces using the spherical cap harmonics (SCH), whose basis functions are a generalisation of the hemispherical harmonics (HSH) basis. SCHs form an orthogonal basis and is suitable for analysing an open surface that is projected onto a spherical cap with any half-angle $\theta_c$ (Figure~\ref{FIG:spherical_cap}), while HSHs form a basis for a hemisphere ($\theta_c=\pi/2$). SCHs have been widely used in geophysics for describing fields (e.g. the magnetic field over a continent), but to our knowledge, they have not yet been applied for characterising and reconstructing the morphology of surfaces. We exploited the solution for the even set of Legendre functions of the first kind together with the Fourier functions to constitute the final SCHs that satisfy the Laplace equation on a spherical cap. We chose the even set to perform surface reconstruction with free edges using the Neumann boundary conditions while simultaneously conducting a power spectral analysis using the orthogonality of the even basis functions. We also used SCH to analyse simply connected surfaces with vertices, edges and triangulated faces (the Standard Triangle Language STL files). The first step of such analyses is to couple each vertex on the surface with a unique pair of $\theta$ and $\phi$ (surface parameterisation). Thus, we proposed a parameterisation algorithm that provides conformal one-to-one mapping with minimal area distortion over a unit spherical cap with a prescribed half-angle $\theta_c$ ($\mathbb{S}^2_{\theta \leq \theta_c}$). 
\par
We showed that the proposed analysis approach is invariant to $\theta_c$ yet is sensitive to the distortion introduced in the parameterised surface at different half angles ($\theta_c$'s). The unavoidable distortion caused by the parameterisation algorithm can limit the analysis to certain half angles. Therefore, we used a global optimisation approach to identify the optimal half-angle $\theta_c$ that minimises the area distortion between the input and parameterised surfaces. As expected, we found that the conventional HSH analysis is suitable for surfaces that can be parameterised over a unit hemisphere without introducing significant area and angular distortions. For nominally flat surfaces, however, the traditional HSH fails to analyse and reconstruct the surfaces correctly because of the distortion.
\par
To show the robustness of the morphological analysis and reconstruction convergence, we applied the proposed method to a scanned face sculpture as a visual benchmark. This benchmark was chosen to contain relatively large and small wavelengths to test the convergence to the smallest details. The face was analysed for two selected half angles to demonstrate that the method is invariant to $\theta_c$ and that the quality of the analysis depends only on the distortion introduced on the parameterised surface. We then demonstrated the convergence and the scalability of the method to expand and reconstruct complex surfaces of a selected local patch on the face. Because the patch was smaller than the entire face, fewer degrees were needed to capture the same details (faster convergence). We finally compared this method with the HSH and found that the reconstruction from HSH was wavy and showed larger error margins, as was expected due to the distortion in the parameterisation.
\par
We derived an approximate formula that links the size of the first degree ellipsoidal cap (FDEC) with the wavelengths as a function of the index $k$. We then used the power spectral analysis of the rotation-invariant shape descriptors that are invariant to translation, rotation, scale and half-angle $\theta_c$ to find the fractal dimension (FD) of the analysed open surfaces. To benchmark this approach, we generated fractal surfaces with the traditional PSD method and computed the Hurst exponent for these surfaces from the SCHA results. The method yielded good estimates of the Hurst exponent (errors between $-7.0$ and $+4.4\%$).
\par
As this method was proposed to study rough surfaces in contact, we also outline a method for projecting real or artificially generated roughness onto a selected donor mesh. This method can be used for the traditional SH, HSH and the SCH. We limited the bandwidth in the spectral domain to reconstruct and alter the roughness of surfaces between user-defined upper and lower limits.
\par
The computational cost of this method was found especially expensive when evaluating the sequential hypergeometric function, the critical loop, for $k > 12$. This is mostly due to using a high-level programming language such as MATLAB and can be bested with using C or C++.

\section{Future works}
\label{SEC:FUTURE}
In this section, we outline future developments for SCHA:
\begin{itemize}
    \item This work hinges on the numerical stability and accuracy of the computed basis functions of SCH, which depends on the Gaussian hypergeometric function evaluation $_2F_1(a,b;c;z)$. We provide a proof-of-concept implementation in MATLAB, which may not be optimal in terms of computational efficiency. We now need a faster implementation of the hypergeometric functions in a more efficient environment (e.g. C++) that is also stable for high orders.
    
    \item The herein proposed method can be seen as a corner stone for regional wavelet analyses for shape morphology. Wavelet analyses conducted via traditional methods like SH could take advantage of the stable and well-studied evaluation techniques of the traditional basis function as well as any traditional recursive formulae (e.g. SH basis). The SCH serves as an exact solution to compare with other potential regional analysis methods using traditional basis functions. 
    
    \item Implementing a recursive approach for computing the SCH basis can improve the speed and efficiency.
    
    \item A physical regularisation method based on optimising the estimated SCH coefficients of the basis will improve the accuracy and noise levels of the analysis (see \cite{REF:8} as an example for SHA).
\end{itemize}

\section{Reproducibility}
The codes and generated data and the surfaces are all made available in this paper and can be found on Zenodo (GitHub) \cite{SCH_codes}:
\begin{itemize}
    \item Zenodo:~\href{https://zenodo.org/record/4890809#.YLZLepozZhE}{10.5281/zenodo.4890809}
    \item GitHub:~\href{https://github.com/eesd-epfl/spherical-cap-harmonics}{git@github.com:eesd-epfl/spherical-cap-harmonics}
\end{itemize}

\section*{Acknowledgements}
We thank Prof. Jean-Fran\c{c}ois Molinari and his group, especially Dr. Guillaume Anciaux, for their valuable comments on the manuscript. We would also thank Prof. Edward And\`{o} for his valuable input and thorough review. {This work was supported in part by the National Science Foundation under Grant No.~DMS-2002103 (to G.P.T.C.).}

\section*{Declaration of Competing Interest}
The authors declare that they have no known competing financial interests or personal relationships that could have appeared to influence the work reported in this paper.

\section{Appendix A: Supplemental results}
\label{SEC:APPX}
\subsection{The scanned stone and the ordinary spherical harmonics analysis}
\label{SUBSEC:APP_LS_STONE}
\par
\begin{figure}
\centering
\includegraphics[width=0.6\textwidth]{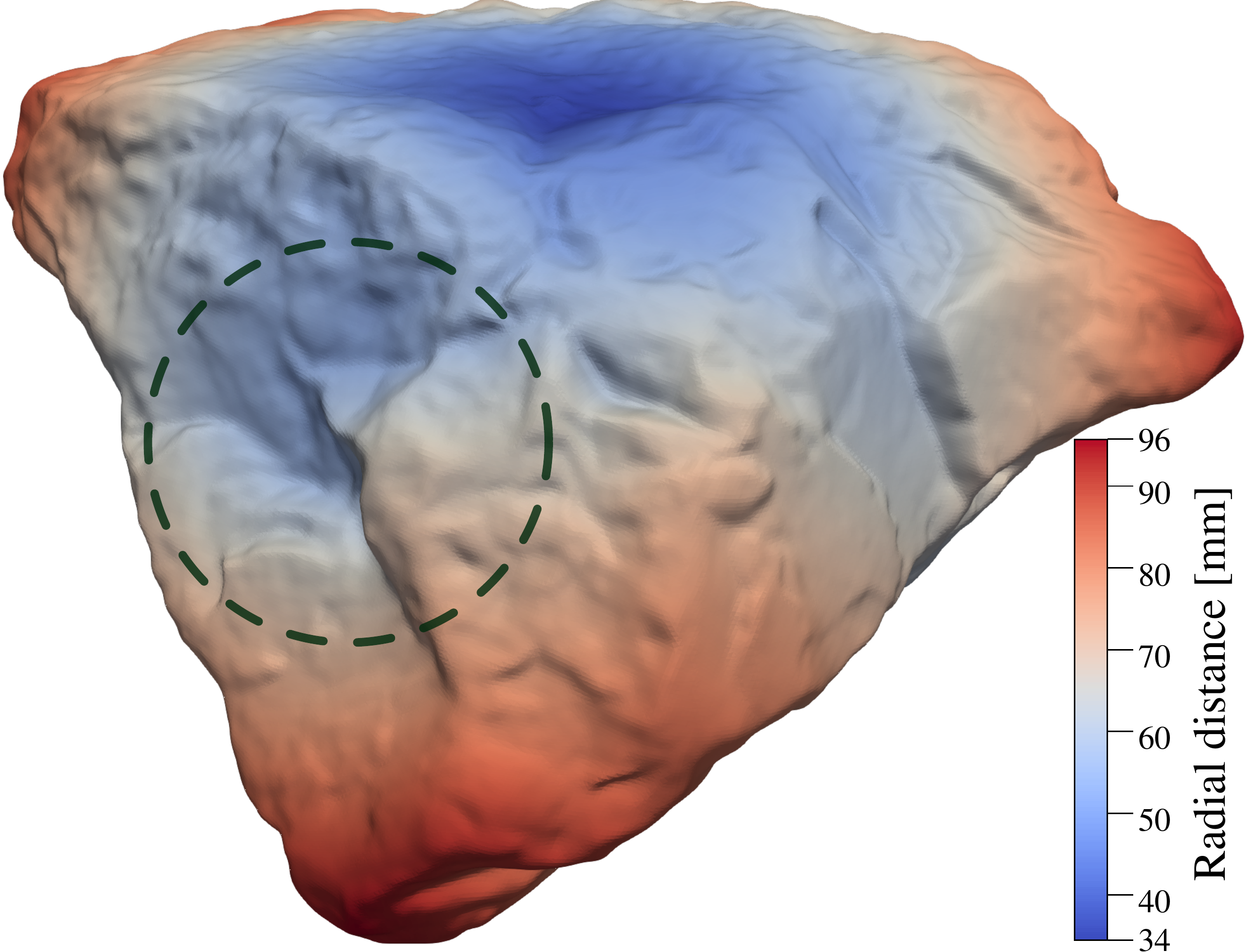}%
\caption{The laser scanning data after down-sampling the point cloud to $444,653$ points instead of nearly $23,000,000$ points; the colour map shows the radial distance measured from the centroid of the stone.
\label{FIG:LS_STONE_RESULTS}
}
\end{figure}
\begin{figure*}
\centering
\includegraphics[width=0.6\textwidth]{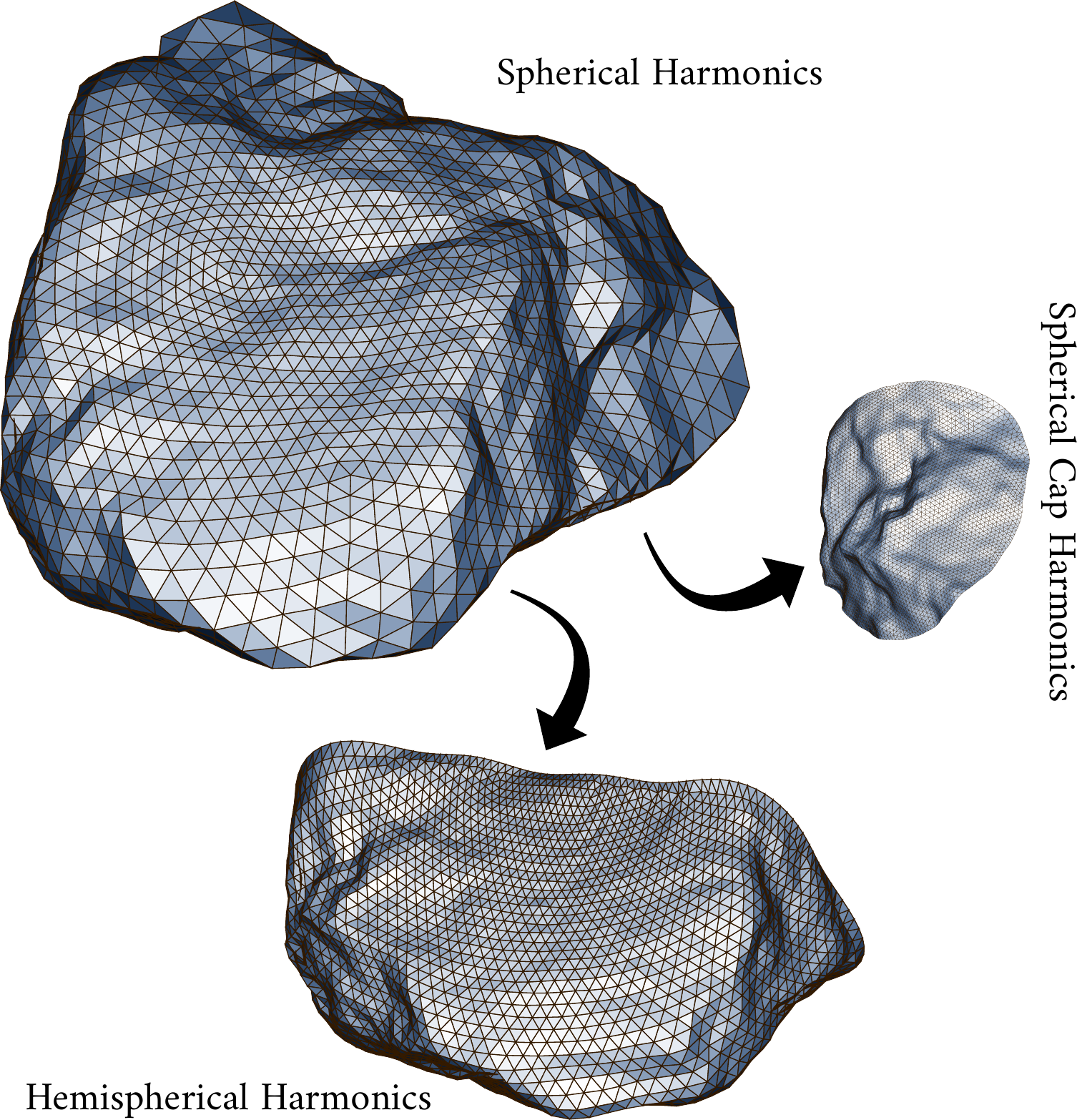}%
\caption{The reconstruction of the stone made using SH to analyse the whole stone ($L_{\max} = 45$), HSH for a regional analysis ($L_{\max} = 45$) and SCH for the regional analysis of nominally-flat patches ($K_{\max} = 40$).
\label{FIG:COMP_SH_HSH_SCH_REC}
}
\end{figure*}

Figure \ref{FIG:LS_STONE_RESULTS} shows the scans of the stone produced using the FARO arm laser scanner after down-sampling the point cloud and remeshing the surface. Using the first degree of the ellipsoid (FDE), which results from the degree $l = 1$ in the SHA, we can estimate the maximum ellipse size at $\theta = \pi/2$. The size of the ellipse can be written as \cite{REF:1}:
\begin{equation}
    \label{EQN:FDE_MAT}
    A = \frac{\sqrt{3}}{2\sqrt{2\pi}} \left({}^{\theta_c}q^{-1}_{1}-{}^{\theta_c}q^{1}_{1}, i({}^{\theta_c}q^{-1}_{1} + {}^{\theta_c}q^{1}_{1}), \sqrt{2}{}^{\theta_c}q^{0}_{1}\right).
\end{equation}
For the SHA, we used $45$ degrees and set the size of the maximum ellipse on the FDE to $a = 66.913$ mm, $b = 65.088$ mm and $c = 47.998$ mm. This corresponds to $\omega_{\max} = 207.3668$ mm (at $l = 1$) and $\omega_{\min} = 4.8225$ mm (at $l = 45$), which was calculated using the approximate formula (\ref{EQN:WAVELENGTH}). The result of the reconstruction of the SH is shown in Figure \ref{FIG:COMP_SH_HSH_SCH_REC}.
\par
The size of the resulted basis function matrix, from SHA, is $(l_{\max} + 1)^2 \times n_{v}$ with exactly $2,116 \times 444,653$, and if we use double precision with $\times 64$ bit-based systems, the size required to store only the basis matrix will be $7.5270$ gigabytes. With the SCH, we captured details up to a wavelength of $\omega_{\min} = 1.99$ mm with only $k = 40$ (equivalent to $l = 40$), making the size of the basis matrix stored only $220.1438$ megabytes. The former calculations were estimated based on the IEEE Standard for Floating-Point Arithmetic (IEEE 754) \cite{REF:64}.

\subsection{Optimal half-angle \texorpdfstring{$\theta_c$}{Lg} for a part of the scanned stone}
\label{SUBSEC:APP_OPT_THETA_C}
\par
\begin{figure}
\centering
\includegraphics[width=0.6\textwidth]{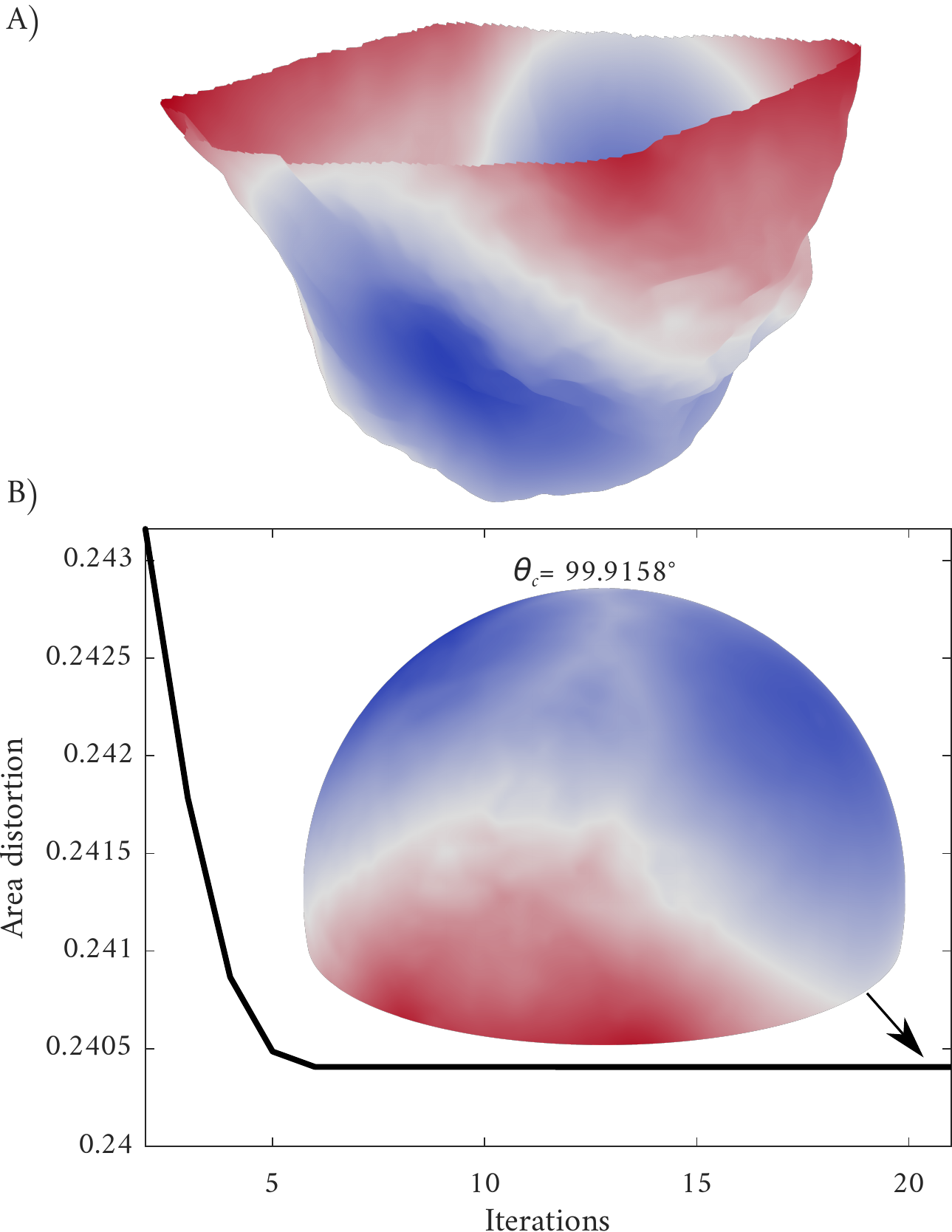}%
\caption{The optimal half-angle $\theta_c$ for the spherical cap parameterisation obtained for the area extracted from the scanned stone in Figure \ref{FIG:LS_STONE_RESULTS}; the map shows the radial distance from the centroid of the extracted part.
\label{FIG:OPT_THETA_C_LS}
}
\end{figure}

Figure \ref{FIG:OPT_THETA_C_LS} shows the convergence to the optimal half-angle $\theta_c$ that minimises the area distortion for extracted part of the scanned stone. The results were obtained by solving Eq.~(\ref{EQN:OPT_TH_C}) using the PSS algorithm. The HSH reconstruction result is shown in Figure \ref{FIG:COMP_SH_HSH_SCH_REC}.

\subsection{Hemispherical harmonics for nominally flat surfaces}
\label{SUBSEC:APP_HSH}
\par
\begin{figure}
\centering
\includegraphics[width=0.5\textwidth]{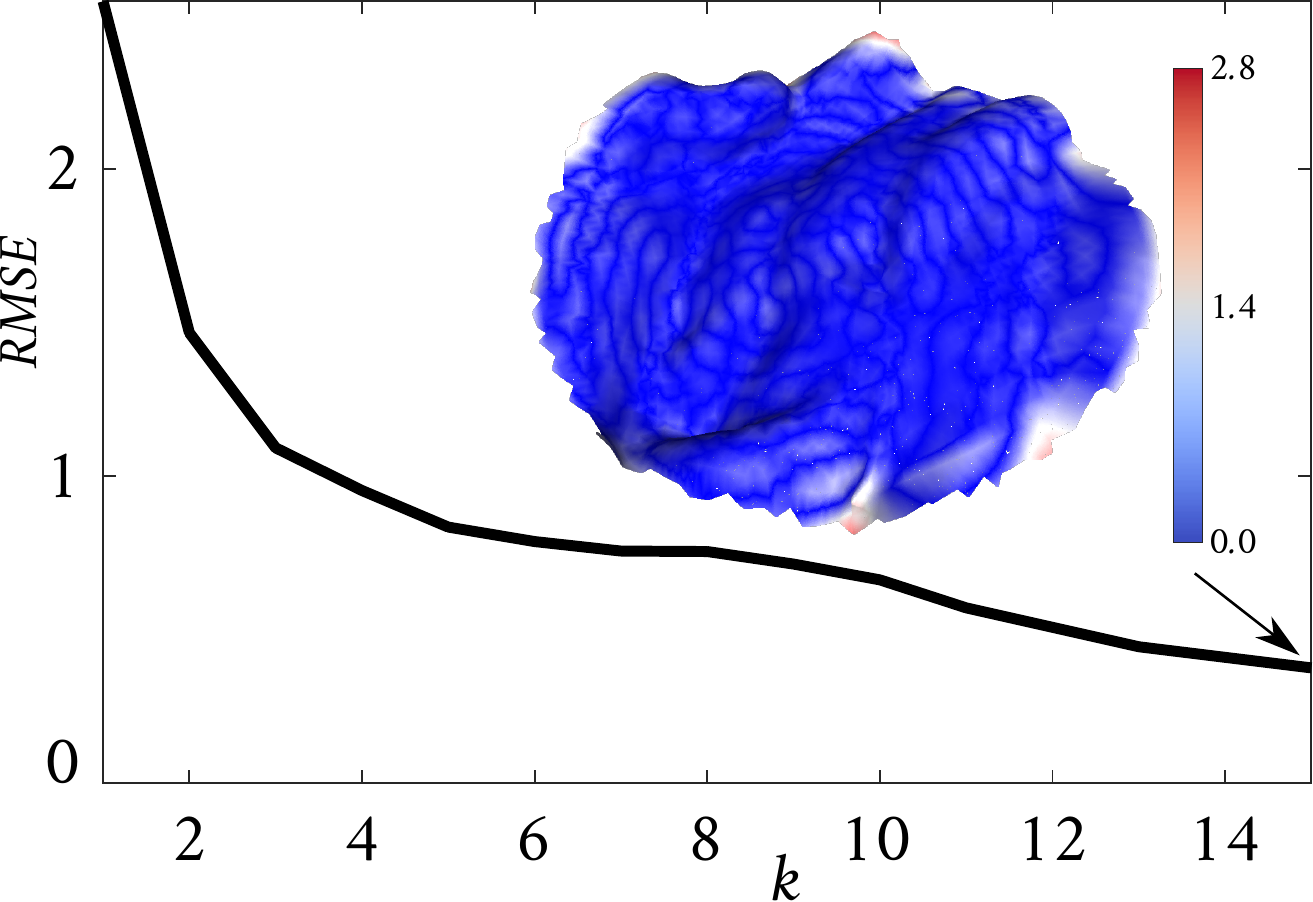}%
\caption{The root mean square error (RMSE) for the local patch on the visual benchmark, as computer using HSH. The inset shows the Hausdorff distance at $k = 15$ where we used $1,202,087$ points sampled on the original mesh and compared this value with the reconstructed mesh. At $\theta_c = \pi/18$, the RMSE for the Hausdorff distance was $0.310431$, and the mean error was $0.087238$, while the highest error value scored was $1.594269$. The reported data was extracted using the open source package MeshLab \protect\cite{REF:47}.
\label{FIG:FACE_PATCH_RES_RMSE_HSH}
}
\end{figure}

Using HSH (refer to the basis functions in \cite{REF:19}) to analyse and reconstruct nominally flat surfaces introduces waviness and does not necessarily require many more degrees to converge as the reconstruction will not improve. To demonstrate this, we analysed the surface patch sampled over the 3D face shown in Figures \ref{FIG:FACE_RES_Ks} and \ref{FIG:FACE_PATCH_RES_RMSE} with HSH. By assuming that $l = 15$ (the same maximum order used in Figure \ref{FIG:FACE_PATCH_RES_RMSE}), we found that the RMSE was $0.283845$ instead of $0.104812$ as found by the SCHA. The RMSE of Hausdorff was $0.310431$ instead of $0.149861$, with a mean value of $0.174521$ compared with $0.087238$ in SCHA. The area distortion introduced by the parameterisation algorithm for a hemisphere was $0.4691203496$ compared to $0.2187252197$ for $\theta_c = \pi/18$. Using orders beyond $l = 15$ in the HSH analysis, the overfitting waviness starts appearing in the reconstruction because of the increasing error from the least-squares regularisation method. Figure \ref{FIG:FACE_PATCH_RES_RMSE_HSH} summarises the analysis results. A closer look at the Hausdorff distance (the inset from the same figure) shows a little waviness in the domain.

\subsection{Additional results for the Sturm-Liouville eigenvalues}
\label{SUBSEC:SL_eign}
\begin{figure}
\centering
\includegraphics[width=0.7\textwidth]{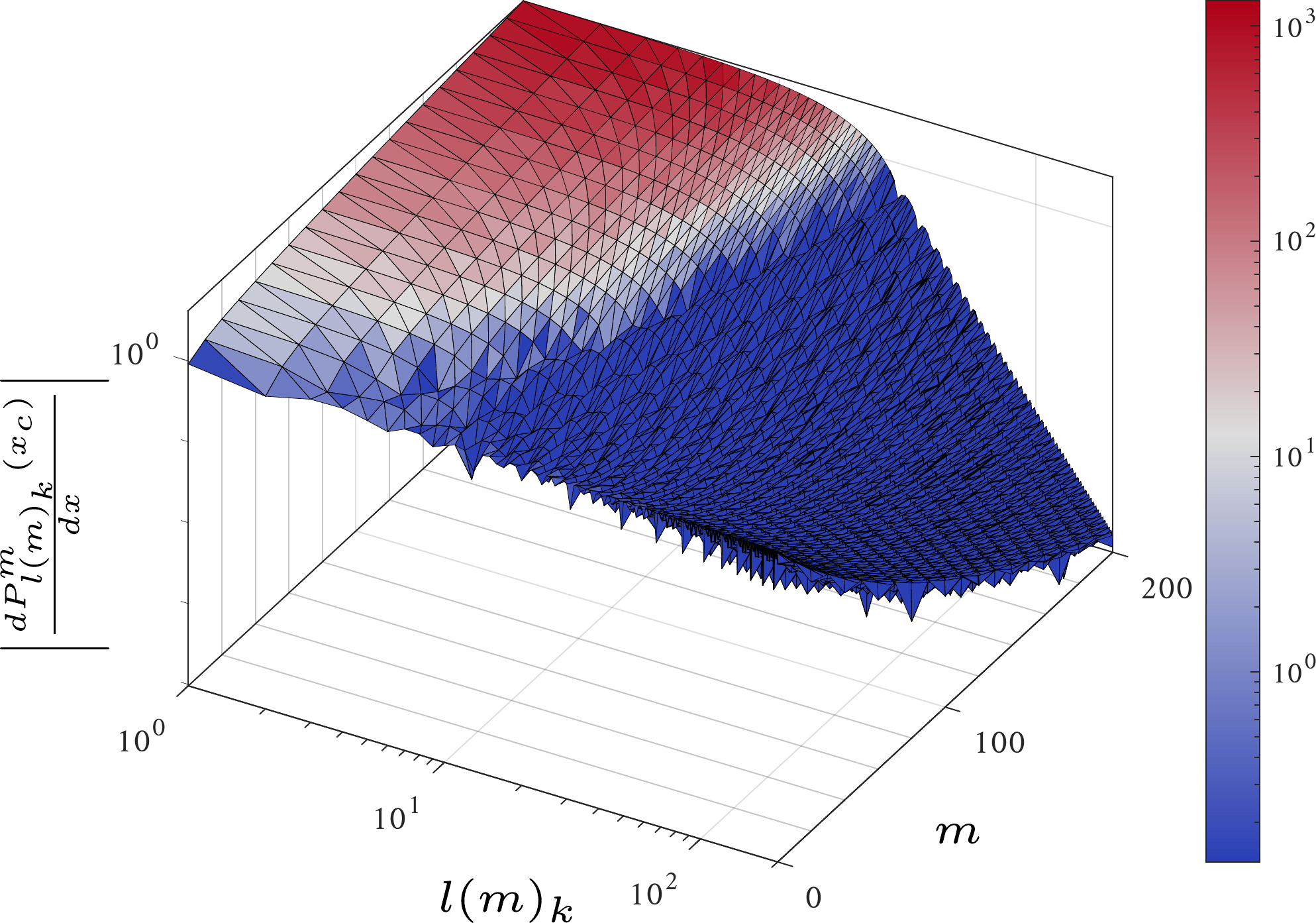}%
\caption{Surface plot of the boundary condition (\ref{EQN:BC}) plotted on a log scale along the $l(m)_k$ and $\bigg|\frac{d P^m_{l(m)_k}(x_c)}{dx}\bigg|$ axes and plotted on the arithmetic scale for $m$ (even Legendre's basis at $\theta_c = 5\pi/18$). The pointy locals on the surface are the locations of the eigenvalues when $\log{\bigg|\frac{d P^m_{l(m)_k}(x_c)}{dx}\bigg|} \to -\infty$.
\label{FIG:EIGN_TH50_surf}
}
\end{figure}

\begin{figure}
\centering
\includegraphics[width=0.6\textwidth]{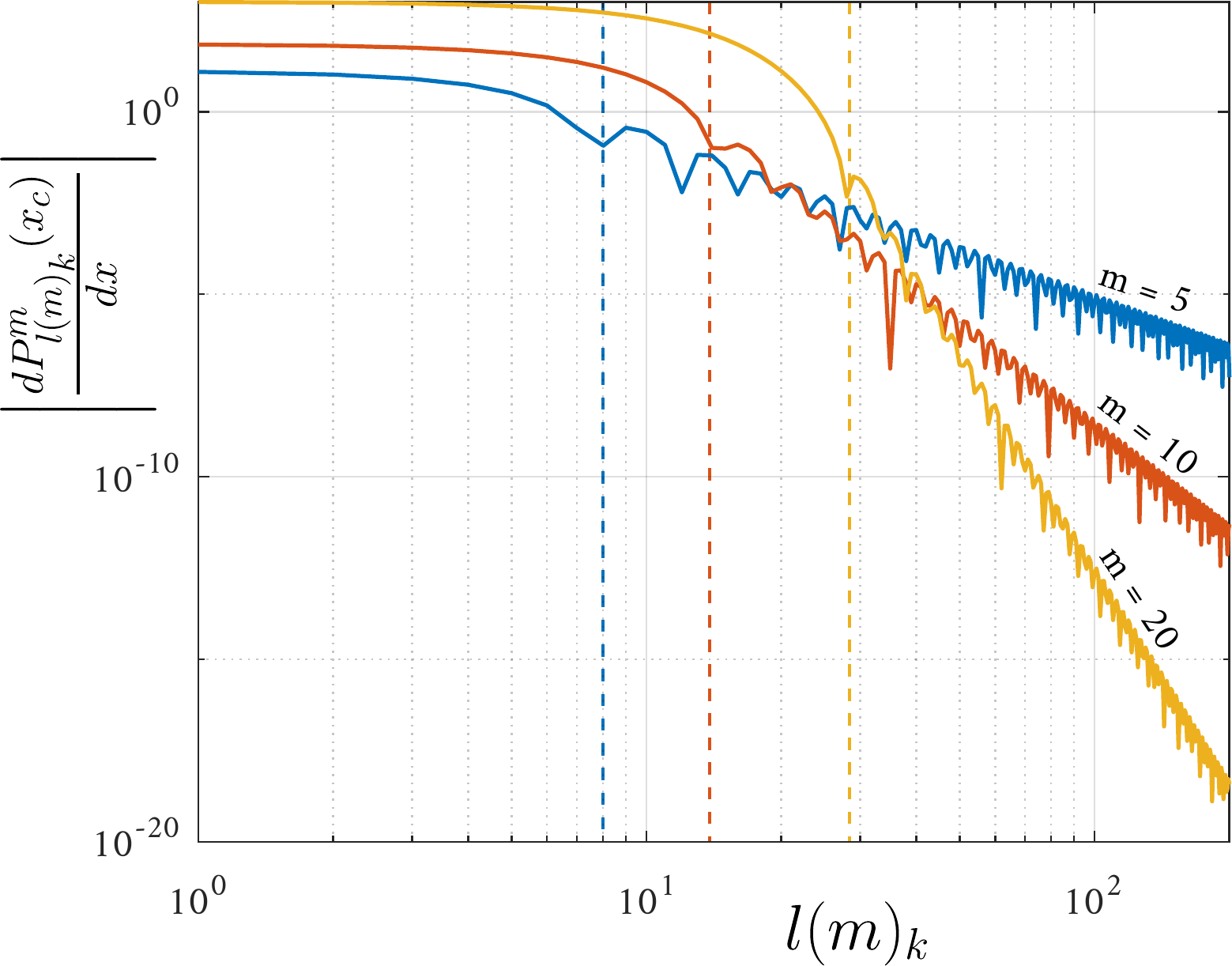}%
\caption{Plot of the boundary condition in Eq.~(\ref{EQN:BC}) for $m \in \{5,10,20\}$ (even Legendre's basis at $\theta_c = 5\pi/18$). The vertical dashed lines mark the first identified eigenvalue at the end of the plateau area and define the size of these plateaus for each $m$. Notice the matched colour code of the solid and dashed lines. The vibrations (local minima) announce the locations of the eigenvalues for different $k$s as $\log{\bigg|\frac{d P^m_{l(m)_k}(x_c)}{dx}\bigg|} \to -\infty$.
\label{FIG:EIGN_TH50_3_orders}
}
\end{figure}

\begin{figure}
\centering
\includegraphics[width=0.7\textwidth]{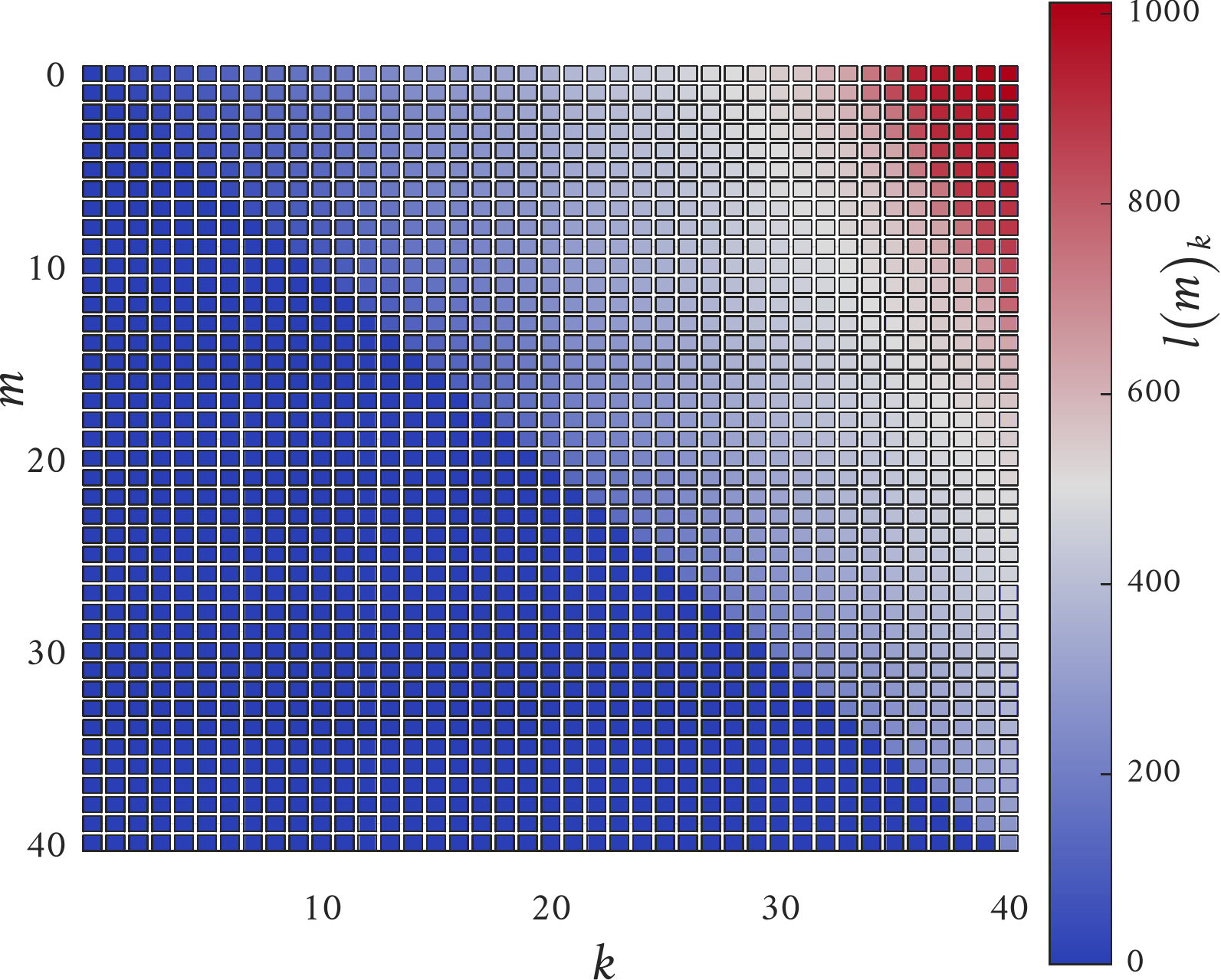}%
\caption{Eigenvalues for the even set when the half-angle $\theta_c = \pi/18$ and for $k \in \{0,1,\dots,40\}$. Notice when $m>k$ $l(m)_k = 0$.
\label{FIG:EIGN_K_40_TH10}
}
\end{figure}

In this section, we show the Neumann boundary conditions and eigenvalues of the S-L problem for visualising the sequence of eigenvalues and their effects on the wavelengths. Figure \ref{FIG:EIGN_TH50_surf} shows the boundary condition of the even set. The locations of the eigenvalues on this figure occur where we see the asymptotic points all over the surface that approach $-\infty$ when the $\frac{d P^m_{l(m)_k}(x_c)}{dx}$ $\to 0$ (eigenvalues), localising the points (asymptotics). These asymptotic lines are more obvious when the equations are plotted with greater accuracy (smaller step-size). Figure \ref{FIG:EIGN_TH50_3_orders} shows the boundary conditions for three selected orders and explains how the size of the plateaus increase with $m$ and thus how they affect the wavelength contributions at different levels when they come into effect in the expansion series. Figure \ref{FIG:EIGN_K_40_TH10} shows an example of the identified roots (eigenvalues) for the even boundary conditions at $\theta_c = \pi/18$.

\bibliography{References}
\label{SEC:REF}

\end{document}